\documentclass[11pt]{article}

\usepackage[margin=1in]{geometry}
\usepackage{multirow}

\usepackage{amsmath,amssymb,amsthm}
\usepackage{graphicx}
\usepackage{booktabs}
\usepackage{hyperref}
\usepackage{natbib}
\usepackage{xcolor}

\newcommand{\E}{\mathbb{E}}

\newcommand{\Prob}{\mathbb{P}}

\newcommand{\cN}{\mathcal{N}}

\usepackage{subcaption}

\usepackage[section]{placeins}

\usepackage{algorithm}
\usepackage{algpseudocode}

\newcommand{\stokesRmsObsOverSigmaMean}{1.35}

\newcommand{\stokesNobs}{200}
\newcommand{\stokesNmcKThree}{100}

\newcommand{\stokesMorozovDetKThree}{1\%}
\newcommand{\stokesFixedChiSqDetKThree}{100\%}
\newcommand{\stokesBatchFourierDetKThree}{100\%}
\newcommand{\stokesEpDetKThree}{100\%}

\newcommand{\stokesEpMedianStopKThree}{19}
\newcommand{\stokesPocockDetKThree}{100\%}

\newcommand{\stokesPocockMedianStopKThree}{26}
\newcommand{\stokesOBFDetKThree}{100\%}

\newcommand{\stokesOBFMedianStopKThree}{54}
\newcommand{\stokesEpMedianStopFractionKThree}{10\%}

\newcommand{\stokesNmcKFour}{100}
\newcommand{\stokesNullMorozov}{0\%}
\newcommand{\stokesNullBatchFourier}{6\%}
\newcommand{\stokesNullEp}{2\%}

\newcommand{\stokesNullPocock}{3\%}
\newcommand{\stokesNullOBF}{13\%}

\newcommand{\stokesThetaTruthFormatted}{$\theta = (-0.60, 1.70, 0.30, 0.50)$}
\newcommand{\stokesThetaOracleKThreeFormatted}{$\hat\theta = (-0.80, 2.24, 0.27)$}
\newcommand{\stokesBetaPointwiseMaxKThree}{71\%}
\newcommand{\stokesBetaPointwiseMeanKThree}{33\%}
\newcommand{\stokesUxRelErrInteriorKThree}{0.07\%}
\newcommand{\stokesBetaRegionalMeanKThree}{42\%}
\newcommand{\stokesBetaArgminTruth}{0.63}
\newcommand{\stokesBetaArgminKThree}{0.52}

\newcommand{\stokesRegionalIntervalLo}{0.25}
\newcommand{\stokesRegionalIntervalHi}{0.75}

\newcommand{\PoisHzeroEPRate}{0.005}
\newcommand{\PoisHzeroBonfRate}{0.085}
\newcommand{\PoisHzeroPocockRate}{0.030}
\newcommand{\PoisHzeroOBFRate}{0.070}
\newcommand{\PoisHzeroMorozovRate}{0.000}
\newcommand{\PoisHzeroFCRate}{0.045}
\newcommand{\PoisHzeroBFRate}{0.035}
\newcommand{\PoisHzeroECount}{1}

\newcommand{\PoisPieceLamPoneEDet}{100}

\newcommand{\PoisPieceLamPoneBonfDet}{39}

\newcommand{\PoisPieceLamPoneBonfMed}{18}

\newcommand{\PoisBumpLamPzeroFiveRMS}{0.0047}
\newcommand{\PoisBumpLamPzeroFiveEDet}{100}

\newcommand{\PoisBumpLamPzeroFiveBonfDet}{52}

\newcommand{\PoisBumpLamPzeroFiveMrzvDet}{0}

\newcommand{\PoisBumpLamPoneBonfMed}{29.5}

\newcommand{\PoisThreeLamPoneEDet}{100}

\newcommand{\PoisThreeLamPoneBonfDet}{98}

\newcommand{\PoisThreeLamPoneFiveBonfMed}{20.5}

\newcommand{\PoisLinearLamPfiveEDet}{100}

\newcommand{\PoisLinearLamPfiveBonfDet}{42}

\newcommand{\PoisLinearLamOneRMS}{0.0057}
\newcommand{\PoisLinearLamOneEDet}{100}

\newcommand{\PoisLinearLamOneEMed}{50}

\newcommand{\PoisLinearLamOneBonfDet}{87}

\newcommand{\PoisLinearLamOneMrzvDet}{0}

\newcommand{\PoisBumpReassureUerr}{12.0}
\newcommand{\PoisBumpReassureAvgErr}{5.4}
\newcommand{\PoisBumpReassureFluxErr}{2.6}

\newcommand{\PoisBumpReassureEMed}{23}

\newcommand{\PoisBumpReassurePocockMed}{37}

\newcommand{\PoisBumpReassureOBFMed}{99.5}

\newcommand{\PoisLinearReassureUerr}{15.3}
\newcommand{\PoisLinearReassureAvgErr}{7.8}
\newcommand{\PoisLinearReassureFluxErr}{3.9}

\newcommand{\PoisLinearReassureEDetRate}{1.0}
\newcommand{\PoisLinearReassureEMed}{44.5}
\newcommand{\PoisLinearReassureEMedPct}{9}

\newcommand{\PoisLinearReassureFCDet}{100}

\newcommand{\PoisLinearReassurePocockDetRate}{0.975}
\newcommand{\PoisLinearReassurePocockMed}{136}

\newcommand{\PoisLinearReassureOBFMed}{192}

\newcommand{\PoisLinearTopThreeFirstCount}{200}
\newcommand{\PoisLinearTopThreeSecondCount}{199}
\newcommand{\PoisLinearTopThreeThirdCount}{105}
\newcommand{\PoisLinearTopThreeNRuns}{200}

\newcommand{\IceERMS}{1.34}
\newcommand{\IceEDet}{100}
\newcommand{\IceEMed}{37}
\newcommand{\IceEMean}{40}

\newcommand{\IceMrzvDet}{1}
\newcommand{\IceMrzvAcceptPct}{99}

\newcommand{\IceFCDet}{100}

\newcommand{\IceBFDet}{100}

\newcommand{\IcePocockDet}{100}
\newcommand{\IcePocockMed}{31.5}
\newcommand{\IcePocockMean}{35}
\newcommand{\IceOBFDet}{100}
\newcommand{\IceOBFMed}{61}
\newcommand{\IceOBFMean}{60}
\newcommand{\IceBonfDet}{100}
\newcommand{\IceBonfMed}{25.5}
\newcommand{\IceBonfMean}{35}
\newcommand{\IceENullRate}{0.010}
\newcommand{\IceMrzvNullRate}{0.000}
\newcommand{\IceBFNullRate}{0.000}
\newcommand{\IcePocockNullRate}{0.010}
\newcommand{\IceOBFNullRate}{0.030}
\newcommand{\IceBonfNullRate}{0.100}

\newcommand{\IceBasalFrictionErrMaxPct}{51}

\newcommand{\IceSurfaceVelErrMedianPct}{1.4}
\newcommand{\IceTerminusFluxErrPct}{36}

\newcommand{\CorrLinearInfFiveUerr}{3.4}

\newcommand{\CorrLinearInfTenUerr}{3.8}

\newcommand{\CorrBumpNoCorrUerr}{12.0}
\newcommand{\CorrBumpNoCorrAvgErr}{5.4}
\newcommand{\CorrBumpInfThreeUerr}{16.9}
\newcommand{\CorrBumpInfThreeAvgErr}{8.0}
\newcommand{\CorrBumpInfFiveUerr}{4.6}

\newcommand{\CorrBumpInfTenUerr}{4.5}

\newcommand{\CorrThreeStepNoCorrUerr}{20.1}

\title{Sequential Structure-Sensitive Residual Diagnostics for PDE Inverse Problems} 
\author{
  Ieva Kazlauskaite\\
  \\
   \small London School of Economics and Political Science\\
}

\date{}

\begin{document}

\maketitle        

\begin{abstract}
Computational models in science and engineering are often assessed by checking whether the residual norm is consistent with the assumed noise level. This can be misleading in smoothing inverse problems: structured model errors may be attenuated in observation space, leaving residual magnitudes below practitioner discrepancy thresholds while coherent residual patterns remain. As a result, residual-norm diagnostics can accept fitted models that still give biased parameters, predictions, or quantities of interest. We propose a structure-sensitive sequential diagnostic based on e-processes. The method uses a portfolio of spatial residual-pattern experts, updates their likelihood-ratio wealth as observations are processed, and rejects the fitted model when the aggregate wealth crosses a prescribed threshold, giving anytime-valid type-I error control for a fixed fitted model. We compare the method with Morozov discrepancy checks, fixed-sample residual tests, and batch projection tests. Across three inverse problems (elliptic diffusion, two-dimensional Stokes flow, and a glaciological ice-stream inversion implemented in the community finite-element model \textit{icepack}) we demonstrate how standard discrepancy checks accept misspecified fits that produce materially wrong quantities of interest. Structure-sensitive batch tests detect these failures using the full dataset, while the e-process detects them earlier from a fraction of the observations. After rejection, the expert wealth attributes the evidence to residual patterns in the chosen dictionary and provides a basis for exploratory model correction.
\end{abstract}

\section{Introduction}

Solving an inverse problem means inferring parameters of a computational model from observations. A separate but essential step is to assess whether the fitted model is adequate for prediction. In particular, one must ask whether the residuals are consistent with measurement noise, or whether they contain systematic structure indicating model misspecification~\citep{kennedy2001bayesian,judd2004indistinguishable,berger2019statistical,duffin2021statistical}. This adequacy question is distinct from parameter estimation, and directly affects the reliability of predictions and quantities of interest derived from the fitted model~\citep{brynjarsdottir2014learning}.

Residual-norm diagnostics are common in inverse problems~\citep{engl1996regularization,vogel2002computational, morozov2012methods}. For example, Morozov's discrepancy principle selects regularisation parameters by matching the residual norm $\|y-G(\hat\theta)\|$ to the noise level $\delta$. It is used for boundary condition and parameter estimation as well as in applications including electrical impedance tomography, seismic waveform inversion, and heat-transfer inverse problems~\citep{jin2010numerical,petra2012inexact, borcea2002electrical,alifanov2012inverse}. Similar logic is also used informally for model adequacy: if the residual magnitude is consistent with the assumed noise level, the fitted model is often treated as acceptable. Chi-squared goodness-of-fit tests on the residuals formalise this idea: under a correctly specified model with Gaussian noise, the sum of squared standardised residuals follows a chi-squared distribution, against which an observed value can be tested.

These magnitude-based diagnostics assess the size of the residuals rather than their spatial structure. This distinction matters in smoothing or otherwise ill-conditioned PDE inverse problems. Structured model error can be attenuated in observation space so that the residual norm remains below a practical discrepancy threshold even though coherent residual patterns remain. A calibrated fixed-sample chi-squared test can detect some such failures, but it requires the completed fit and the full dataset. It also gives limited diagnostic information about the spatial form of the misspecification. A complementary diagnostic should therefore test directly for systematic residual patterns.

In this work we turn structure-sensitive residual testing into a sequential post-fit diagnostic for PDE inverse problems. Given a fitted parameter $\hat\theta$, we process residuals against the fixed model $G_{\hat\theta}$ and compare them with a dictionary of spatial residual patterns. Each pattern defines a likelihood-ratio expert, and the experts are combined into a portfolio e-process. When the aggregate wealth of this portfolio crosses a prescribed threshold, the diagnostic signals structured lack of fit with anytime-valid type-I (false-positive) error control. This gives a practical workflow for detecting residual structure during data acquisition, while also attributing the evidence to the patterns that drive the rejection.

The methodology is based on e-processes, sequential tests from game-theoretic statistics~\citep{ramdas2025hypothesis,vovk2021values}. Their safe-testing interpretation provides type-I error control under optional stopping and optional continuation~\citep{grunwald2024safe,grunwald2024beyond,ter2020anytime,henzi2022valid}. This is useful in experimental and observational settings where diagnostics may be inspected repeatedly, and where additional measurements may be collected after preliminary evidence of lack of fit. The same anytime-valid logic underlies sequential monitoring of deployed predictive models, tracking model risk to flag harmful distribution shifts under continuous inspection~\citep{johari2022always,podkopaev2022tracking}. E-processes are a natural fit for PDE residual diagnostics because the experts are dictated by the geometry of the problem rather than chosen ad hoc: the spatial domain supplies a canonical set of smooth basis functions (\emph{e.g.} Fourier and polynomial modes), and the smoothing action of the forward operator suggests which of these the data can resolve. In this paper we use e-processes in a post-fit regime: the fitted model and noise level are treated as fixed when residual monitoring begins. Fully online refitting of $\hat\theta$ would require additional validity safeguards, such as sample splitting or universal inference~\citep{wasserman2020universal}, and is outside the present scope.

We compare the proposed diagnostic with both magnitude-based and structure-sensitive baselines. Morozov and chi-squared tests represent residual-magnitude diagnostics. A batch Fourier projection test represents a non-sequential structure-sensitive diagnostic applied after all data have been collected. This comparison separates two effects: the benefit of testing residual structure, and the additional benefit of sequential anytime-valid monitoring. The e-process is not intended to replace all batch diagnostics. Its contribution is to provide early detection from a fraction of the observations and an interpretable wealth distribution over residual-pattern experts. The dominant experts do not uniquely identify the physical source of the model error, but they indicate which residual patterns in the chosen dictionary explain the accumulated evidence and can guide follow-up checks or exploratory correction.

The paper makes three contributions. First, we formulate a portfolio e-process for residual-pattern detection that gives anytime-valid early warning for a fixed fitted model. Second, we show how the portfolio weights provide interpretable attribution to spatial residual patterns and can suggest parsimonious follow-up corrections. Third, we empirically demonstrate in three PDE inverse problems that discrepancy-threshold diagnostics can accept misspecified models whose residuals are small but structured, while the resulting quantities of interest remain biased.

\section{Method} \label{sec:methods}

\subsection{Problem setup}\label{sec:setup}
Consider a PDE-based inverse problem in regression form~\citep{stuart2010inverse}. A forward operator $G_\theta$ maps parameters $\theta$ to the solution $u$ of the PDE, and observations are taken at spatial locations $X_t$,
\begin{equation}\label{eq:obs}
  Y_t = g_{\mathrm{true}}(X_t) + \epsilon_t, \quad \epsilon_t \sim \cN(0, \sigma^2), \quad t = 1, 2, \ldots,
\end{equation}
with the locations fixed or chosen independently of the observation errors, and $g_{\mathrm{true}}$ the true observation mean. The model is correctly specified if $g_{\mathrm{true}} = G_\theta$ for some $\theta \in \Theta$, and misspecified otherwise; the cases we study are deliberately the latter.

After fitting $\hat\theta$, we fix the forward model $G_{\hat\theta}$ and test the residuals against it. The residual-monitoring null is
\begin{equation}\label{eq:null}
  H_0(\hat\theta): \quad r_t = Y_t - G_{\hat\theta}(X_t) \sim \cN(0, \sigma^2)\ \text{independently},
\end{equation}
with the observation locations $X_t$ fixed before each residual is observed. This is a \emph{conditional, post-fit} adequacy null: it asks whether the residuals of the fixed fit are consistent with pure noise, not whether the model class $\{G_\theta : \theta \in \Theta\}$ contains the truth. Even a correctly specified class can leave a nonzero residual mean if the fit is imperfect (for example, an optimiser that stops at a local minimum); and when the truth lies outside the class, no $\theta$ removes the residual structure at all. It is this structure that the diagnostic targets, whatever its source. The martingale and anytime-valid type-I guarantees of \S\ref{sec:method-portfolio} hold conditionally on $\hat\theta$ and $\sigma$.

The null~\eqref{eq:null} holds exactly for monitoring observations that are independent of the fit, such as new or held-out data tested against a fixed $G_{\hat\theta}$. When the same noisy observations are used both to estimate $\hat\theta$ and to test, the residuals are coupled through the fit and~\eqref{eq:null} becomes an approximation; preserving exact validity then requires sample splitting, held-out data, or universal inference~\citep{wasserman2020universal}.

Under misspecification, the residuals of the fitted model have structured spatial mean,
\begin{equation}\label{eq:residual_alt}
  r_t \sim \cN(\mu^*(X_t),\, \sigma^2), \quad \mu^*(x) = g_{\mathrm{true}}(x) - G_{\hat\theta}(x),
\end{equation}
the observation-space discrepancy between the true mean and the fitted model. If the truth lies in the model class, $g_{\mathrm{true}} = G_\theta$ and this reduces to $\mu^*(x) = G_\theta(x) - G_{\hat\theta}(x)$.

\subsection{Background: e-values and e-processes}\label{sec:method-eprocesses}

Our diagnostic is built from \emph{e-values} and \emph{e-processes}, statistical objects from game-theoretic hypothesis testing~\citep{ramdas2023game,ramdas2025hypothesis,grunwald2024safe,vovk2021values}. We summarise the main concepts; for a comprehensive overview, see~\citet{ramdas2025hypothesis}.

\paragraph{E-values.} Fix a null hypothesis $H_0$, understood as a set $\mathcal{P}$ of candidate distributions for the data $Y$. An \emph{e-variable} is a nonnegative statistic $E = E(Y)$ whose expectation is at most one under every distribution in the null, $\E_P[E] \leq 1$ for all $P \in \mathcal{P}$; the number it returns on the observed data is the corresponding \emph{e-value} (as is common, we let ``e-value'' refer to the e-variable itself where no confusion arises). This expectation constraint is the defining contrast with a p-value, which is constrained in probability instead, and it is what makes $E$ usable as a test on its own: by Markov's inequality $\Prob_P(E \geq 1/\alpha) \leq \alpha$, so rejecting $H_0$ when $E \geq 1/\alpha$ is a level-$\alpha$ test, and a larger e-value is stronger evidence against the null.

E-variables generalise likelihood ratios. When the null is simple, $H_0 = \{P_0\}$, the likelihood ratio $E = q(Y)/p_0(Y)$ against any other distribution $Q$ is an e-variable, since $\E_{P_0}[E] = \int q(y)\,dy = 1$; conversely, every e-variable that attains $\E_{P_0}[E] = 1$ is the likelihood ratio against some distribution $Q$, so on the boundary the two notions coincide. For composite nulls e-variables still exist but need not be ordinary likelihood ratios, and are best read as generalisations of them~\citep[\S3]{ramdas2025hypothesis}. The freedom in the choice of $Q$ is a modelling freedom: taking $Q$ to be a mixture $q(Y) = \int q_\theta(Y)\, w(\theta)\, d\theta$ over a family of alternatives $\{Q_\theta\}$ with weights $w$ makes $E = q(Y)/p_0(Y)$ a Bayes factor, which against a simple null is again an e-variable. Our residual-noise null is simple, so each expert in the bank below is a genuine likelihood ratio and the uniform mixture~\eqref{eq:eprocess} is exactly such a Bayes-factor e-variable, constructed in \S\ref{sec:method-portfolio}.

It is useful to read an e-value as the payoff of a bet against the null with unit stake~\citep{shafer2021testing}. Specifically, the bet is fair or unfavourable whenever $H_0$ holds, so no strategy can make money in expectation under the null; sustained gains are therefore evidence against it. Each expert is one such bet on a specific spatial residual pattern, and its accumulated wealth measures how well that pattern explains the observed residuals. Under misspecification, an expert aligned with the true residual structure makes systematic gains, and its wealth grows without bound as evidence accumulates; the mixture inherits this growth and eventually crosses the rejection threshold.

\paragraph{E-processes.} Now let monitoring observations arrive one at a time. At step $t$ the location $X_t$ is fixed in advance, or chosen from past observations only, and then the residual $r_t$ is observed; let $E_t$ be a nonnegative statistic computed from the first $t$ residuals. 
The sequence $(E_t)_{t \geq 1}$ is an \emph{e-process} if it remains an e-value at every (almost surely finite) data-dependent stopping time: for any rule $\tau$ that decides whether to stop using only the data seen so far,
\begin{equation}\label{eq:eprocess-def}
  \E_P[E_\tau] \leq 1 \qquad \text{for all } P \in \mathcal{P}.
\end{equation}
For the threshold-crossing rule we use, validity holds for the full observation budget and follows directly from Ville's inequality (\S\ref{sec:method-portfolio}).

The practical content of~\eqref{eq:eprocess-def} is that the stopping rule need not be fixed, or even known, in advance. One may stop at a preset sample size, after a suspicious run of residuals, or as soon as the evidence is decisive (``stop at the first $t$ with $E_t \geq 1/\alpha$''), and $E_\tau$ is still a valid e-value. This is the property classical tests lack: a p-value is valid only at a sample size fixed \emph{ahead} of time, and Wald's sequential test requires its stopping boundary to be specified in advance~\citep{ramdas2025hypothesis}. With an e-process the analyst may inspect the diagnostic as often as desired, and continue or halt data collection in response to what is seen, without inflating the type-I error. The same logic allows \emph{optional continuation}: a later round of observations can be folded into the running e-value by multiplication, provided each new factor is itself an e-value conditional on the earlier data; this is the case when the new factor is built from fresh observations using a betting rule fixed in advance of seeing them. Under this condition the product remains valid even when the decision to gather more data was prompted by the earlier evidence~\citep{grunwald2024safe}. This is the basis for the continuous monitoring use case. Combining~\eqref{eq:eprocess-def} with Markov's inequality gives the level-$\alpha$ rule we use: reject $H_0$ at the first $t$ with $E_t \geq 1/\alpha$.

One way to construct an e-process is a test martingale: a nonnegative martingale started at $E_0 = 1$~\citep[\S7]{ramdas2025hypothesis}. For these, Ville's inequality~\citep{ville1939etude},
\begin{equation}\label{eq:ville}
  \Prob_{H_0}\!\left(\sup_{t \geq 1} E_t \geq \frac{1}{\alpha}\right) \leq \alpha,
\end{equation}
bounds the probability that the trajectory \emph{ever} crosses $1/\alpha$, which is exactly the type-I error of the rule above. Our diagnostic is of this form: the likelihood-ratio mixture $E_t$ in~\eqref{eq:eprocess} is a nonnegative martingale started at one under the residual-noise null, and so inherits the calibration~\eqref{eq:ville} directly.

\subsection{Expert construction}\label{sec:experts}

The e-process of \S\ref{sec:method-eprocesses} requires a family of alternatives for the experts to bet on. Under misspecification the residuals are not pure noise: they carry the deterministic mean $\mu^*$ of~\eqref{eq:residual_alt}, the structure the fit has missed expressed in observation space. Each expert proposes one candidate shape for this mean through the residual model $r_t \sim \cN(\mu_k(X_t), \sigma^2)$, and accrues wealth through its likelihood ratio against the noise-only null $r_t \sim \cN(0, \sigma^2)$ when the residuals carry that pattern. The bank $\{\mu_k\}$ is the dictionary of shapes that the experts test for, and the design question is which shapes to include.

A principled choice follows from how model error typically looks. For instance, in the model discrepancy framework of~\citet{kennedy2001bayesian}, the mismatch between model and reality is itself modelled as a smooth Gaussian process, $\delta(x) \sim \mathrm{GP}(0, k_\ell)$, and the residual mean $\mu^*$ is exactly the realisation of such a discrepancy as seen through the forward operator. Smooth GP discrepancies admit Karhunen--Lo\`eve expansions in low-frequency bases: for a squared-exponential kernel on $[0,1]$ with Lebesgue measure, the leading eigenfunctions are well approximated by Fourier modes $\{\sin(j\pi x), \cos(j\pi x)\}$ (the precise eigenfunctions depend on the domain, boundary conditions, and kernel). This motivates a smooth, low-frequency basis for the bank; its effectiveness across our examples is empirical rather than canonical.

PDE problems are particularly well-suited to expert-based testing: the spatial domain comes with a canonical inner product, smooth bases (Fourier, Chebyshev, polynomial) span discrepancies that respect the regularity of the forward operator, and the smoothing properties of the operator suggest which modes the data can resolve. A generic regression problem on unstructured features would require ad-hoc basis design or an adaptive scheme; for PDE residuals, the geometry suggests natural low-frequency bases. The cost of including extra experts is logarithmic in $K$~(\S\ref{sec:method-portfolio}), so the practitioner can be generous with the bank without much loss in detection power.

Specifically, we draw the experts from a Fourier--polynomial basis
\begin{equation}\label{eq:experts}
  \{\mu_k\} = \bigl\{a \sin(j\pi x),\; a\cos(j\pi x)\bigr\}_{j=1}^J \cup \bigl\{a(x\!-\!\tfrac{1}{2})^d\bigr\}_{d=1}^3
\end{equation}
at 12 signed amplitude levels $a \in \{\pm 0.5\sigma, \pm 1\sigma, \pm 2\sigma, \pm 3\sigma, \pm 5\sigma, \pm 8\sigma\}$. Each likelihood-ratio expert is one-sided, testing a residual mean aligned with $+a\phi$; including both signs ensures the bank can detect a discrepancy of either polarity, since the $+a\phi$ expert has negative expected log-likelihood growth against $-a\phi$ and would otherwise miss it. The amplitude magnitudes discretise the unknown discrepancy scale, playing the role of the prior on the GP variance in the Bayesian framework. With $J = 5$ frequencies ($5 \times 2 = 10$ sine/cosine shapes) and 3 polynomial shapes, the 13 spatial patterns at 12 signed amplitude levels yield $K = 156$ experts.

\subsection{Universal portfolio e-process}\label{sec:method-portfolio}
 
We construct $E_t$ by the standard method of mixtures for composite alternatives~\cite[\S3.7]{ramdas2025hypothesis} (see also Cover's universal portfolio~\citep{cover1991universal}). At time $t$, the instantaneous log-likelihood ratio for expert $k$ is
\begin{equation}\label{eq:lr}
  \ell_{t,k} = \log\frac{\cN(r_t;\, \mu_k(X_t),\, \sigma^2)}{\cN(r_t;\, 0,\, \sigma^2)} = \frac{2\,r_t\,\mu_k(X_t) - \mu_k(X_t)^2}{2\sigma^2}.
\end{equation}
The cumulative log-likelihood ratio for expert $k$ after $t$ observations is $L_{t,k} = \sum_{i=1}^t \ell_{i,k}$. The e-process is the mixture
\begin{equation}\label{eq:eprocess}
  E_t = \sum_{k=1}^K \pi_k \exp(L_{t,k}),
\end{equation}
with uniform prior $\pi_k = 1/K$, and we reject when $E_t \geq 1/\alpha$ at significance level $\alpha$ (\emph{e.g.} $E_t \geq 1/\alpha = 20$ at significance level $\alpha = 0.05$). 
Under $H_0$, and conditional on the fitted $\hat\theta$ and a predictable observation design (each location is fixed before its residual is observed), the residuals are independent $\cN(0,\sigma^2)$. Because the location $X_t$ is fixed before $r_t$ is observed, the proposed mean $\mu_k(X_t)$ is non-random given the past. Under the null $r_t \sim \cN(0,\sigma^2)$, so a direct calculation with the Gaussian moment generating function gives $\E[\exp(\ell_{t,k}) \mid \text{past}] = 1$ under the null, so $\E[\exp(L_{t,k})\mid\text{past}] = \exp(L_{t-1,k})$: each $\exp(L_{t,k})$ is a nonnegative martingale started at $1$. A convex combination of martingales is again one, so $E_t$ is a test martingale and hence an e-process.

\paragraph{Order invariance and stopping time.} The terminal statistic $E_T$ is a sum of commutative cumulative log-likelihood ratios and is therefore independent of the order in which observations are processed. The rejection decision we actually use is sequential: $H_0$ is rejected if $E_t \geq 1/\alpha$ at any $t \leq T$. This crossing-time rule does depend on the observation sequence, both through the stopping time $\tau = \inf\{t : E_t \geq 1/\alpha\}$ and through the binary outcome, since whether the running maximum $\sup_{t \leq T} E_t$ reaches $1/\alpha$ can itself depend on the order in which increments arrive. Ville's inequality nonetheless applies to the crossing-time rule for any predictable ordering and bounds the type-I error at $\alpha$ uniformly over orderings, so validity does not depend on the choice of order. Our Poisson and Stokes experiments use independent random permutations within each run, while the \textit{icepack} experiment uses one fixed cached observation order. Both schemes are predictable because the ordering is chosen without examining future residuals, and Appendix~\ref{sec:si-ordering} quantifies the empirical sensitivity of detection rate and stopping time to ordering.
 
Under the alternative~\eqref{eq:residual_alt}, the expected per-step log-likelihood ratio for expert $k$ is $\E[\ell_{t,k}] = \mu_k(X_t)(2\mu^*(X_t) - \mu_k(X_t))/(2\sigma^2)$, maximised over functions at $\mu_k = \mu^*$; a perfectly matched expert earns $\mu^*(X_t)^2/(2\sigma^2)$ at each step, accumulating evidence at a rate set by the size of the model error relative to the noise. The mixture~\eqref{eq:eprocess} satisfies the lower bound
\begin{equation}\label{eq:regret}
  \log E_t = \log\!\left(\sum_k \pi_k \exp(L_{t,k})\right) \geq \max_k L_{t,k} - \log K,
\end{equation}
a deterministic consequence of $\sum_k \pi_k a_k \geq \max_k \pi_k a_k$ with $a_k = \exp(L_{t,k})$ and $\pi_k = 1/K$. The mixture's log-evidence therefore trails the best single expert's by at most $\log K$ at every $t$; that implies a fixed price, logarithmic in the bank size, for mixing over experts rather than knowing the best one in advance.

The complete procedure (Algorithm \ref{alg:eprocess}) is a post-processing step on the residuals of a \emph{fixed} fitted model: given $G_{\hat\theta}$, the update at each observation is $O(K)$ arithmetic operations, negligible relative to a single PDE solve. It requires no modification to the forward solver, no Monte Carlo steps, and no retraining. The guarantees hold when the fit is independent of the monitored residuals, as required by the conditional null~\eqref{eq:null}. This is the case when $G_{\hat\theta}$ is fixed before monitoring begins, using an independent or pre-existing fit, or when the monitored data are collected separately from the data used to fit. When the same observations are used both to fit $\hat\theta$ and to test, the residuals are coupled through the fit (in a linear-Gaussian analogy, $r = (I-H)\epsilon$, with correlated, heteroscedastic entries), and the procedure is then an approximation requiring sample splitting, cross-fitting, or universal inference for exact validity. Our Poisson and Stokes experiments use fits independent of the Monte Carlo noise, and \S\ref{sec:robustness} shows that noisy refitting changes their results little. The \textit{icepack} experiment instead evaluates the full same-data fit-and-test pipeline; its reported type-I rates are therefore empirical and do not directly inherit the conditional fixed-$\hat\theta$ martingale guarantee.

\begin{algorithm}[t]
\caption{E-process for model adequacy monitoring}
\label{alg:eprocess}
\begin{algorithmic}[1]
\Require Significance level $\alpha$, noise $\sigma$, experts $\{\mu_k\}_{k=1}^K$, frozen fit $G_{\hat\theta}$
\State \textbf{Input:} a fit $G_{\hat\theta}$ obtained independently of the monitored data (an independent/training fit, a pre-existing fit, or the oracle fit of \S\ref{sec:robustness}); freeze $G_{\hat\theta}$, $\sigma$, and the expert bank
\State Initialise $L_{0,k} \gets 0$ for all $k$,\ \ $E_0 \gets 1$
\For{monitoring observations $(X_t, Y_t)$, $t = 1, 2, \ldots$, not used to construct $\hat\theta$}
    \State Compute residual $r_t \gets Y_t - G_{\hat\theta}(X_t)$
    \For{$k = 1, \ldots, K$}
    \State $L_{t,k} \gets L_{t-1,k} + \dfrac{2\,r_t\,\mu_k(X_t) - \mu_k(X_t)^2}{2\sigma^2}$
\EndFor
    \State $E_t \gets \displaystyle\sum_{k=1}^K \frac{1}{K}\exp(L_{t,k})$
    \If{$E_t \geq 1/\alpha$}
        \State \textbf{reject} $H_0$
    \EndIf
\EndFor
\end{algorithmic}
\end{algorithm}

The same per-step drift governs the mixture's ability to detect the misspecification. Since $E_t \geq \pi_k \exp(L_{t,k})$ for every expert, it is enough that one expert has positive asymptotic drift. Writing $\langle f, g\rangle_\nu = \int fg\,d\nu$ for the inner product under $\nu$, the density of observation locations (\emph{e.g.} uniform), this drift is
\begin{equation}
  \gamma_k = \frac{2\langle\mu_k, \mu^*\rangle_\nu - \|\mu_k\|_\nu^2}{2\sigma^2} > 0,
\end{equation}
that is, the expert $\mu_k$ is positively aligned with the true discrepancy $\mu^*$ and not so large that its own-size penalty $\|\mu_k\|_\nu^2$ outweighs the alignment reward $2\langle\mu_k,\mu^*\rangle_\nu$. Writing $r_i = \mu^*(X_i) + \epsilon_i$, the cumulative log-likelihood $L_{t,k}$ splits into a deterministic drift converging to $\gamma_k t$ and a martingale-difference noise term; since the bank is finite and bounded on the compact domain, the noise term divided by $t$ vanishes by the law of large numbers, so $L_{t,k}/t \to \gamma_k > 0$ and $L_{t,k} \to +\infty$. This single expert eventually carries the mixture past any fixed threshold. Crucially, the practitioner need not know which expert this is: the mixture detects any misspecification some expert is aligned with, which suits residual diagnostics where the form of the misspecification is unknown. The flip side is that an error orthogonal to every expert goes undetected, so the bank must span the modes the misspecification can excite (\S\ref{sec:bank-sensitivity}). This is the trade-off safe testing is built around~\citep{grunwald2024safe}: a mixture forgoes optimality against any single alternative for validity and power across the whole class.

Two variants of the portfolio e-process are worth noting. The uniform prior over experts could be replaced by a continuous one, integrated out analytically to give a closed-form mixture leading to a standard method-of-mixtures construction in sequential likelihood-ratio testing. We keep the discrete, uniformly weighted experts because they remain individually interpretable: each can be ranked and inspected in the diagnosis step, and the mixing cost is logarithmic in their number regardless. Alternatively, an adaptive portfolio that reweights experts at each step~\citep{waudby2025universal} would compete with the best fixed convex combination of experts rather than the best single one, again at the cost of a less interpretable wealth distribution.
 
\subsection{Comparators}

The e-process combines two properties, sensitivity to the spatial \emph{structure} of the residuals and \emph{sequential} monitoring, and a fair assessment should separate their contributions. We therefore compare it against diagnostics that differ in two respects: whether they test residual magnitude or residual structure, and whether they operate in batch or sequentially. 

The e-process is both structure-sensitive and sequential while each comparator gives up at least one of these properties, so the contribution of each can be isolated. Morozov's discrepancy principle is included separately as a practical reference point. The batch Fourier projection test is the primary structure-sensitive comparator: it uses the same expert bank as the e-process but applies a single Bonferroni-corrected $z$-test after all observations, isolating the contribution of sequential monitoring from that of structure sensitivity. Concretely, for each \emph{unique spatial shape} $\phi_m$ in the bank (the amplitude and sign copies of a given shape produce $Z$ statistics equal up to sign and collapse to a single $|Z_m|$ test), compute $Z_m = \sum_t r_t \phi_m(x_t) / (\sigma\sqrt{\sum_t \phi_m(x_t)^2})$, which is $\cN(0,1)$ under the fixed-fit Gaussian null. We reject if $\max_m |Z_m| > z_{1-\alpha/(2M)}$, the $1-\alpha/(2M)$ quantile of the standard normal, where $M = 13$ is the number of unique shapes in the default bank; the union bound over the $M$ shapes controls the familywise error at $\alpha$. The fixed-sample $\chi^2$ test is the batch magnitude-based counterpart, computing $S = \sum_t (r_t/\sigma)^2$ after all observations and rejecting if $S$ exceeds the upper-$\alpha$ quantile of $\chi^2(n)$.

For sequential magnitude-based comparators we include two commonly used group-sequential $\alpha$-spending rules~\citep{jennison1999group}. These allocate a false-alarm budget across repeated looks at the data; they differ in how the budget is distributed over time. Pocock~\citep{pocock1977group} spends it uniformly, applying the same threshold at every look, while O'Brien--Fleming~\citep{obrien1979multiple} (OBF) spends little early and most late. The Pocock and OBF boundaries are derived assuming the interim test statistics are jointly Gaussian, with the specific correlation structure that arises when a normal mean is monitored over accumulating data. Our sequential statistic is instead a cumulative sum of squared standardised residuals, $S_t = \sum_{i\le t}(r_i/\sigma)^2$, which is $\chi^2$-distributed rather than Gaussian and accumulates differently across looks. The published boundaries therefore do not match its joint law, and applying them is a heuristic rather than an exact level-$\alpha$ procedure; we report the resulting empirical type-I rates in Appendix~\ref{sec:si-calibration} and the stopping-time comparisons should be read against those rates.
We also include a Bonferroni-style rule that rejects when $p_t \leq \alpha/t$. Despite the name, this is not a valid Bonferroni allocation: the per-look levels sum to $\sum_{t=1}^{T} \alpha/t \approx \alpha \ln T > \alpha$, so it overspends the budget regardless of dependence. It is correspondingly anti-conservative and tends to reject at very small $t$, where the threshold $\alpha/1 = \alpha$ imposes no multiplicity penalty at all.

Morozov's discrepancy principle (flag if $\mathrm{RMS}(r) > \tau\sigma$ with $\tau = 1.5$, a pragmatic value; sensitivity to $\tau$ is examined in \S\ref{sec:linear-subnoise}) is included as a reference point because of its utility in practice, though it is designed for regularisation parameter selection rather than goodness-of-fit testing and is therefore expected to have low power against structured alternatives~\citep{morozov2012methods}. A tighter calibration of the sequential magnitude test would replace the spending boundaries with anytime-valid time-uniform boundaries~\citep{howard2021time}, but the underlying statistic remains magnitude-based and so inherits the insensitivity to spatial structure that we identify for Morozov and the fixed-sample $\chi^2$ test.

\subsection{Monte Carlo design}\label{sec:method-mc}

Detection rates and stopping times are Monte Carlo estimates over independent noise realisations; quantity-of-interest errors are deterministic for fixed-fit model comparisons and Monte Carlo averages when computed after noisy correction fits. The run counts vary by experiment. The Poisson power curves use 200 runs under $H_0$ ($\lambda = 0$) and 100 runs for combinations of $\lambda \in \{0, 0.02, 0.05, 0.08, 0.1, 0.15, 0.2, 0.3, 0.5, 1.0\}$ and the four misspecification types. The Poisson diagnostic comparison uses 200 oracle-fit runs each for linear $\lambda = 1.0$, bump $\lambda = 0.10$, and three-step $\lambda = 0.15$; the oracle-vs.-noisy robustness table separately checks refitting on noisy data. The Stokes diagnostic comparison uses 100 runs each for the misspecified ($K = 3$) and correct ($K = 4$) fits with a fixed parameter estimate across runs; validity of this oracle shortcut is checked in \S\ref{sec:robustness}. The \textit{icepack} diagnostic comparison uses the full fit-and-test pipeline: for each of 100 noisy data realisations the misspecified $K = 3$ model is refit by Nelder--Mead, and the null calibration is constructed analogously from 100 noisy refits of the correctly specified $K = 4$ model. In the Poisson and Stokes sequential experiments, observations within each run are processed in a random per-run permutation; the \textit{icepack} comparison uses one randomly sampled set of observation locations and the fixed cached observation order. Sensitivity to ordering is examined in Appendix~\ref{sec:si-ordering}.
Detection rates are Monte Carlo estimates: with 100--200 runs the 95\% confidence half-width is roughly $\pm 0.06$--$0.08$ near $0.8$, so small differences in the transition regions should not be over-interpreted.

\section{Numerical experiments}\label{sec:experiments}

We apply the method to three inverse problems where a smoothing forward operator hides structured model error below a residual-magnitude threshold. The three span a controlled synthetic diffusivity problem, an ill-posed 2D boundary inference, and a community ice-flow model. The first is a 1D Poisson diffusivity inversion (\S\ref{sec:poisson}), a controlled setting in which we can dial the misspecification continuously, sweep its spatial structure across different shapes, and map detection power against a known ground truth. The second is a 2D Stokes basal-drag inversion (\S\ref{sec:stokes}), an ill-posed boundary-inference problem in which a smoothing operator carries error from the base to the surface observations. The third is an ice-stream basal-friction inversion in the community model \textit{icepack} (\S\ref{sec:icepack})\footnote{Code for Poisson and Robin examples is available at \url{https://github.com/IevaKazlauskaite/structured-error-eprocess}; the \textit{icepack} examples are available upon request.}.

Across all three we examine the following: Does a residual-magnitude check (Morozov, fixed-sample $\chi^2$) accept the misspecified fit? Do the structure-sensitive tests detect it, and how early does the sequential e-process detect it relative to the batch tests? And does the accepted fit actually mislead, \emph{i.e.}, are the resulting quantities of interest materially wrong? The Poisson problem additionally serves to characterise detection power and calibration in detail; the Stokes and \textit{icepack} problems test whether the conclusions survive in more realistic operators.
 
\subsection{Poisson diffusivity}\label{sec:poisson}
 
We consider the 1D Poisson equation $-(\theta(x)u')' = f(x)$ on $[0,1]$ with homogeneous Dirichlet boundary conditions $u(0) = u(1) = 0$, constant forcing $f = 10$, and noise $\sigma = 0.01$. The null model parameterises diffusivity as $\theta(x) = \exp(q_1 x + q_2 x^3)$ with true $q_1 = 1.1$, $q_2 = 1.5$. The domain is discretised with 500 piecewise-linear finite elements (giving $n=501$ mesh nodes). Since the problem is small, the forward solves use a dense direct solver. Best-fit null parameters are obtained by Nelder--Mead minimisation of \(\sum_t (u_\lambda(x_t)-G_q(x_t))^2\), with a maximum of 500 iterations. With only 2 parameters, convergence is reliable; the oracle robustness check (\S\ref{sec:robustness}, Table~\ref{tab:oracle}) confirms that the fit quality does not affect results. In each Monte Carlo run, the solution is observed once at every node with independent Gaussian noise, and the e-process processes these $n$ observations sequentially in random order reflecting settings such as observations from a sensor network or satellite snapshots; sensitivity to ordering is examined in Appendix~\ref{sec:si-ordering}. The diagnostic uses the default bank of \S\ref{sec:experts}; at this noise level its twelve signed amplitude levels are $a \in \pm\{0.005, 0.01, 0.02, 0.03, 0.05, 0.08\}$, spanning $0.5\sigma$ to $8\sigma$ in magnitude.

\subsubsection{Interpolated misspecification}
 
\begin{figure}
  \centering
  \includegraphics[width=0.8\textwidth]{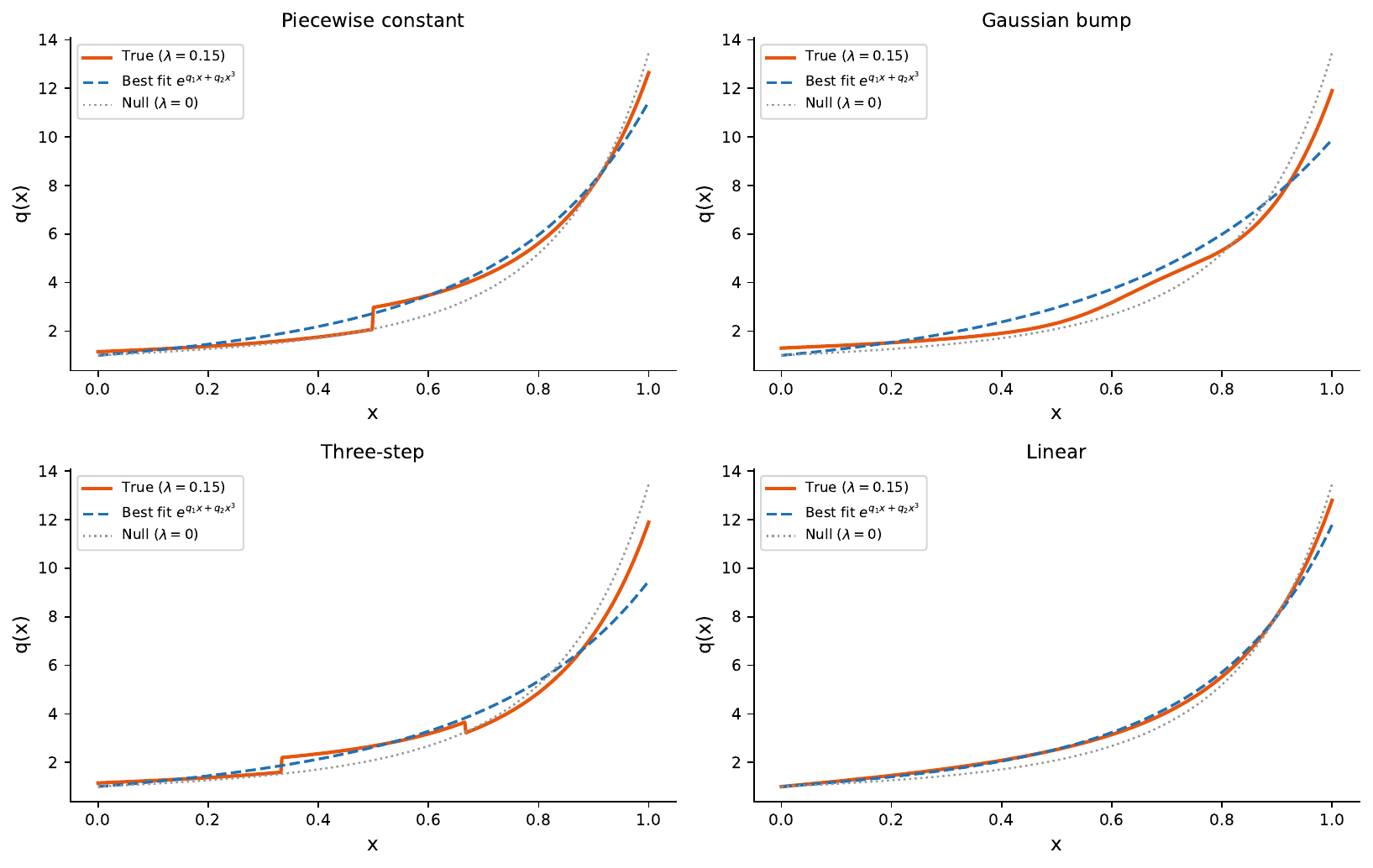}
  \caption{Four misspecification types at $\lambda = 0.15$. True interpolated diffusivity (orange) vs.\ best-fit exponential (blue dashed) and null (gray dotted). The linear case (bottom right, RMS $= 0.00186$) is nearly indistinguishable; the exponential family approximates it well, leaving minimal residual structure. The bump and three-step cases produce larger mismatches.}
  \label{fig:diffusivity}
\end{figure}

We define the interpolated diffusivity $\theta_\lambda(x) = (1-\lambda)\theta_{\mathrm{null}}(x) + \lambda\,\theta_{\mathrm{alt}}(x)$ for $\lambda \in [0,1]$, where $\theta_{\mathrm{alt}}$ takes one of four forms not in the exponential family: piecewise constant, Gaussian bump, three-step, and linear (see Fig.~\ref{fig:diffusivity} and Appendix~\ref{sec:si-geometry} for definitions). These span a range of spatial structures from localised to diffuse. They also span a range of magnitudes relative to the noise: as $\lambda$ increases the model error $\mu^*$ grows from well below the noise level into the super-noise regime for the bump, piecewise, and three-step cases, whereas for the linear case it stays below the noise level at every $\lambda$ ($\|\mu^*\| \leq 0.57\sigma$ even at $\lambda = 1$). This scaling determines whether a residual-\emph{magnitude} check can detect the error, and the answer differs by test. A Morozov-type per-observation RMS threshold gains power asymptotically only when $\|\mu^*\|$ lifts the limiting residual RMS, $\sqrt{\sigma^2 + \|\mu^*\|^2}$, above $\tau\sigma$. The subnoise linear case ($\|\mu^*\| \leq 0.57\sigma$) has limiting RMS below the threshold at $\tau = 1.5$, so increasing the sample size gives the rule no asymptotic power against it beyond its false-alarm rate; a finite sample crosses only through noise fluctuations (\S\ref{sec:linear-subnoise}). An accumulating magnitude test such as the fixed-sample $\chi^2$ behaves differently: its noncentrality $\Lambda = \sum_t (\mu^*(x_t)/\sigma)^2 = n\|\mu^*\|^2/\sigma^2$ grows with $n$, so it eventually detects even a subnoise mean shift, at the cost of requiring the full sample (\S\ref{sec:linear-subnoise}, Table~\ref{tab:reassurance}).

\subsubsection{Type-I error control}\label{sec:type1}
 
\begin{figure}
  \centering
  \includegraphics[width=0.8\columnwidth]{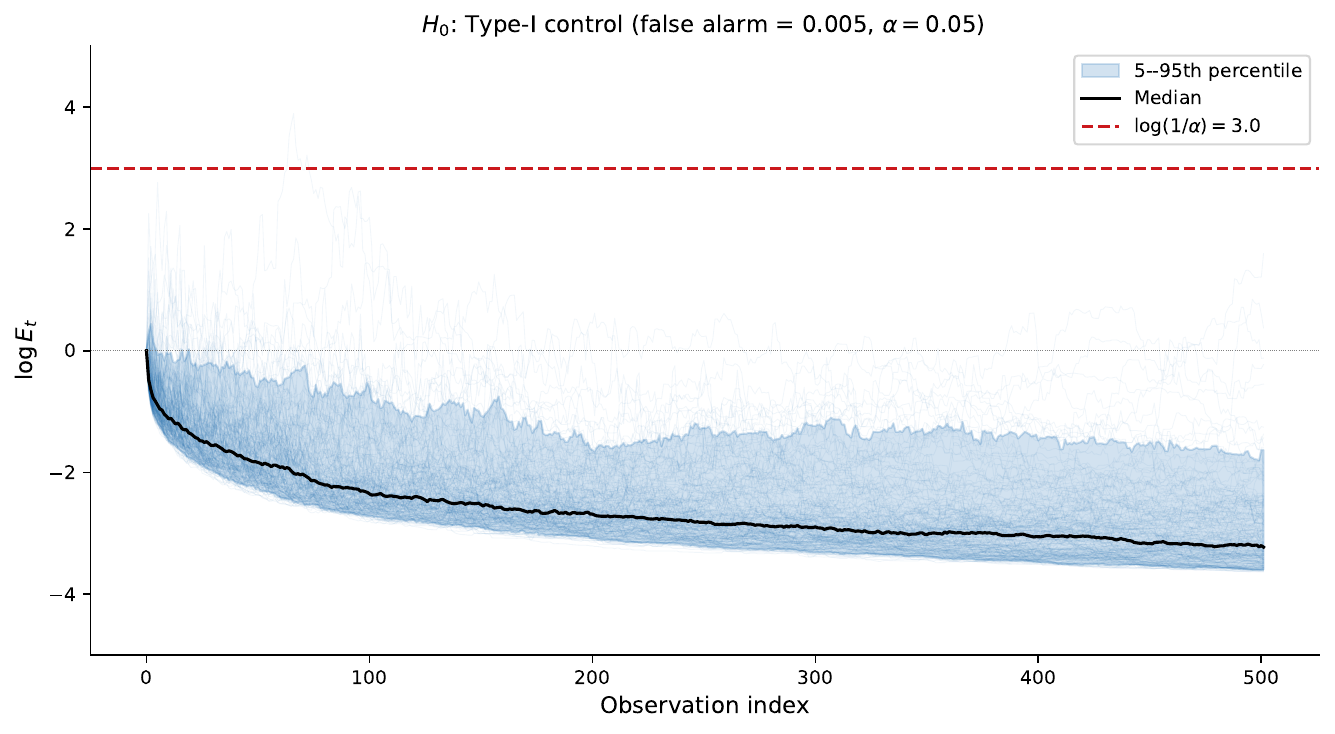}
  \caption{E-process under $H_0$ (correct model). 200 trajectories (light blue), median (black). The 5th--95th percentile band stays well below $\log(1/\alpha) = 3.0$ (red dashed); \PoisHzeroECount{} of $200$ trajectories cross at some point, yielding empirical type-I rate $\PoisHzeroEPRate$. The median drifts to $\approx -3$: expert bets systematically lose against pure noise.}
  \label{fig:type1}
\end{figure}

Under $H_0$ ($\lambda = 0$, correct model, 200 runs), the empirical false alarm rate is $\PoisHzeroEPRate$ (\PoisHzeroECount{} of $200$ trajectories cross the rejection threshold $\log(1/\alpha) = 3.0$ at some point along their trajectory), well below the nominal $\alpha = 0.05$ guaranteed by construction. The median trajectory drifts to approximately $-3$ (Fig.~\ref{fig:type1}), consistent with the theoretical prediction: under $H_0$ expert log-wealths have negative drift under pure noise (even though the wealths themselves remain mean-one martingales), no systematic residual pattern exists, and the 5th--95th percentile band stays well below the threshold throughout, with only the upper tail of trajectories occasionally crossing.

The Bonferroni sequential $\chi^2$ test, by contrast, has an empirical false alarm rate of $\PoisHzeroBonfRate$, exceeding the nominal $\alpha = 0.05$. This inflation is expected: the rule $p_t \leq \alpha/t$ is not a valid Bonferroni allocation, because the per-look levels sum to $\sum_{t=1}^{T}\alpha/t \approx \alpha\ln T > \alpha$, so it overspends its budget regardless of dependence. The small-$t$ boundary of the $\alpha/t$ rule additionally shapes Bonferroni's rejection-time distribution under structured misspecification: across the bump, linear, and three-step functions, $17$--$29\%$ of Bonferroni's rejections occur at $t < 5$ (Appendix~\ref{sec:si-calibration}). The Pocock and OBF spending tests are closer to nominal alternatives with empirical rates $\PoisHzeroPocockRate$ and $\PoisHzeroOBFRate$ respectively. Morozov has a $\PoisHzeroMorozovRate$ false alarm rate. Bonferroni's detection rates in Table~\ref{tab:detection} should be read against its inflated false-alarm rate ($\PoisHzeroBonfRate$): the same boundary effect that lifts its power also lifts its type-I error, so it is not operating at the same level as the other tests. Specifically, since the $\alpha/t$ boundary imposes no multiplicity penalty at $t = 1$, a run can reject on its first observation by chance, and such boundary-admitted rejections pull the median over detecting runs down to very small $t$ (piecewise $\lambda = 0.10$: $\PoisPieceLamPoneBonfMed$ observations). The small medians therefore reflect the rejection rule, not faster genuine detection. The structure-sensitive comparison in Table~\ref{tab:reassurance} excludes Bonferroni for this reason.
 
\subsubsection{Detection power}
 
\begin{figure}
  \centering
  \includegraphics[width=0.8\textwidth]{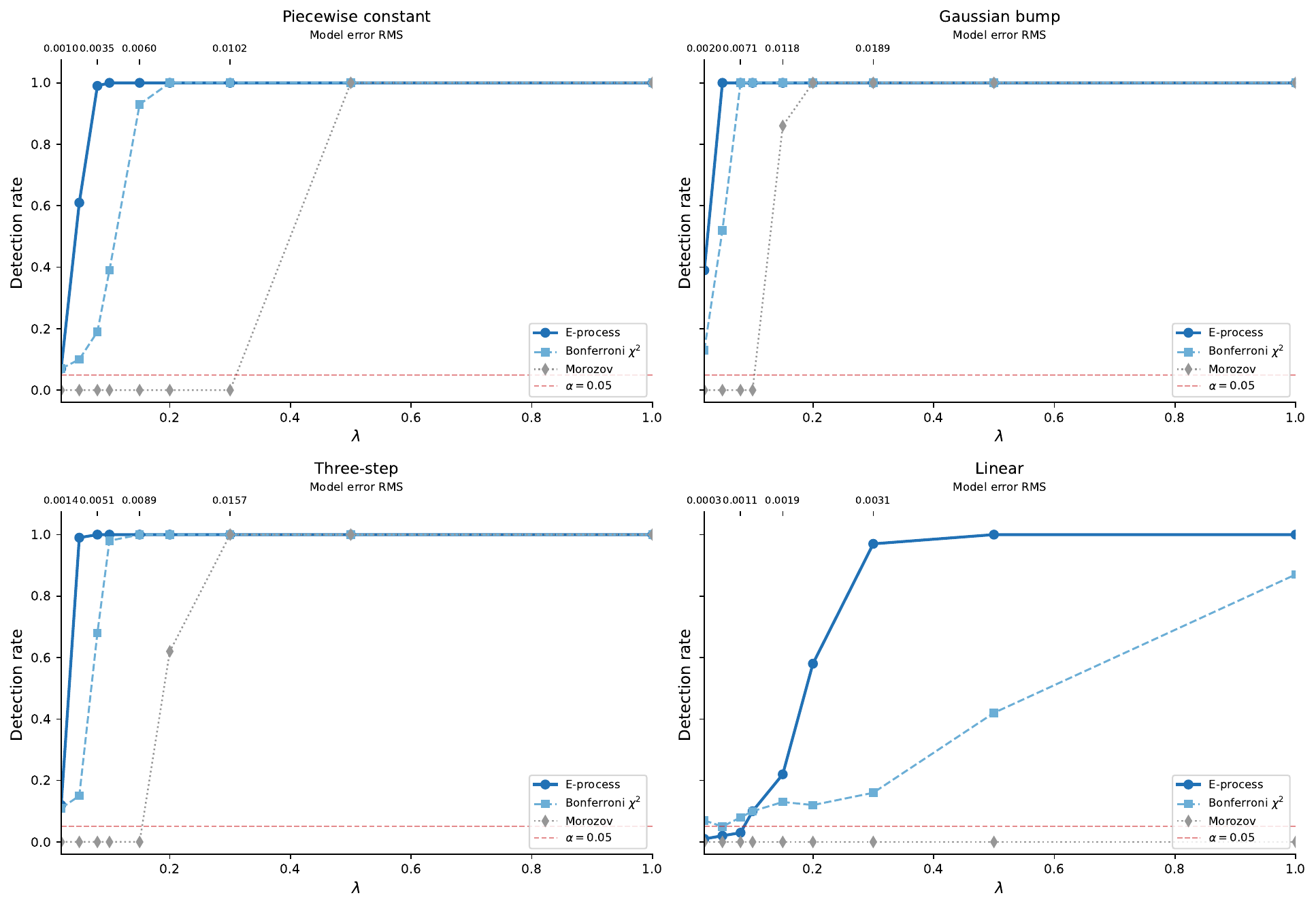}
  \caption{Detection rate vs.\ $\lambda$. E-process (blue), Bonferroni (light blue), Morozov (gray). Top axes: model error magnitude $\|\mu^*\|$. Dashed red: $\alpha = 0.05$. }
  \label{fig:power}
\end{figure}

Fig.~\ref{fig:power} presents detection rates as a function of $\lambda$ across all four misspecification types (100 runs per $\lambda$). Three regimes emerge.
 
For localised misspecification (Gaussian bump), the model error concentrates in a few Fourier modes. The e-process achieves $\PoisBumpLamPzeroFiveEDet\%$ detection at $\lambda = 0.05$ ($\|\mu^*\| = \PoisBumpLamPzeroFiveRMS$, well below $\sigma = 0.01$), compared to $\PoisBumpLamPzeroFiveBonfDet\%$ for Bonferroni and $\PoisBumpLamPzeroFiveMrzvDet\%$ for Morozov.
For distributed misspecification (piecewise, three-step), the pattern is similar. At $\lambda = 0.10$ for piecewise, the e-process detects in $\PoisPieceLamPoneEDet\%$ of runs vs.\ $\PoisPieceLamPoneBonfDet\%$ for Bonferroni. For three-step misspecification, the e-process also dominates in the low-to-moderate transition region; by $\lambda = 0.10$, Bonferroni nearly matches it ($\PoisThreeLamPoneBonfDet\%$ versus $\PoisThreeLamPoneEDet\%$). The structure-sensitive tests and the accumulating fixed-sample $\chi^2$ saturate at nearby larger $\lambda$, while Morozov remains insensitive until the residual RMS itself grows large enough to cross its threshold. The spread across several harmonics makes this alternative less favourable to a single-expert mixture than the concentrated bump case, but the signed $K = 156$ bank still retains equal or higher detection power than Bonferroni throughout the sweep.

For smooth misspecification (linear), the exponential family approximates $1 + 8x$ so well that $\|\mu^*\|$ never exceeds $\PoisLinearLamOneRMS$, even at $\lambda = 1$. Morozov achieves $\PoisLinearLamOneMrzvDet\%$ detection at every $\lambda$. The e-process reaches $\PoisLinearLamPfiveEDet\%$ at $\lambda = 0.5$ and $\PoisLinearLamOneEDet\%$ at $\lambda = 1.0$; Bonferroni reaches $\PoisLinearLamPfiveBonfDet\%$ and $\PoisLinearLamOneBonfDet\%$ respectively. Detection difficulty tracks spatial concentration of model error (see Appendix~\ref{sec:si-geometry}).
 
\begin{table}[htbp]
\centering
\caption{Selected detection rates and median stopping times (conditional on detection, default bank $K = 156$). $H_0$ false alarms: E 0.005, Bonf 0.085, Mrzv 0.000 ($\alpha = 0.05$).}
\label{tab:detection}
\small
\begin{tabular}{@{}llcccrr@{}}
\toprule
 & & \multicolumn{3}{c}{Detection} & \multicolumn{2}{c}{Stop time} \\
\cmidrule(lr){3-5}\cmidrule(lr){6-7}
Type & $\lambda$ & E & Bonf & Mrzv & E & Bonf \\
\midrule
Piecewise constant & 0.05 & 0.61 & 0.10 & 0.00 & 299 & 1 \\
 & 0.10 & 1.0 & 0.39 & 0.00 & 96 & 18 \\
Gaussian bump & 0.02 & 0.39 & 0.13 & 0.00 & 194 & 2 \\
 & 0.05 & 1.0 & 0.52 & 0.00 & 77 & 86 \\
Three-step & 0.05 & 0.99 & 0.15 & 0.00 & 147 & 5 \\
 & 0.08 & 1.0 & 0.68 & 0.00 & 50 & 172 \\
Linear & 0.20 & 0.58 & 0.12 & 0.00 & 220 & 2 \\
 & 0.50 & 1.0 & 0.42 & 0.00 & 90 & 104 \\
\bottomrule
\end{tabular}
\end{table}

\subsubsection{Linear case: detection below the noise floor}\label{sec:linear-subnoise}

The linear case shows the failure mode most clearly. Morozov's rule is a per-observation check on the residual RMS, $\mathrm{RMS}(r) = \sqrt{n^{-1}\sum_{t=1}^n r_t^2}$. Under misspecification the expected mean square of the residuals is $\E[n^{-1}\sum_t r_t^2] = \sigma^2 + \|\mu^*\|^2$, where $\|\mu^*\| = (n^{-1}\sum_{t=1}^n \mu^*(x_t)^2)^{1/2}$ is the root-mean-square magnitude of the model error, and the empirical residual RMS converges to $\sqrt{\sigma^2 + \|\mu^*\|^2}$ as $n$ grows. Its limiting value stays below the threshold $\tau\sigma$ whenever $\|\mu^*\|/\sigma < \sqrt{\tau^2 - 1}$, a region in which increasing the sample size gives the magnitude rule no asymptotic sensitivity to the model error. At $\lambda = 1$ the linear model error has $\|\mu^*\| = \PoisLinearLamOneRMS$, just over $\sigma/2$, placing it inside this region at the standard $\tau = 1.5$ (limiting residual RMS $\approx 0.0115 < \tau\sigma = 0.015$): Morozov has no asymptotic power against this alternative beyond noise-driven false alarms, and in our Monte Carlo runs it accepts in every case ($\PoisHzeroMorozovRate$ false-alarm rate under $H_0$). Tightening $\tau$ toward $1$ shrinks the insensitive region, at the cost of the false-alarm rate.

For the calibrated fixed-sample $\chi^2$ test, its statistic $S = \sum_t (r_t/\sigma)^2$ is central $\chi^2_n$ under the noise-only null but noncentral under misspecification, with noncentrality $\Lambda = \sum_t (\mu^*(x_t)/\sigma)^2$ measuring how far the signal shifts $S$ above its null mean. Unlike the per-observation RMS that Morozov checks, this noncentrality \emph{accumulates} across observations. Against the null distribution of $S$, the shift is more than five null standard deviations higher; the alternative distribution therefore lies almost entirely past the threshold, and the test rejects with probability near one  (Table~\ref{tab:reassurance} confirms $\PoisLinearReassureFCDet\%$ detection). The cost is that this detection arrives only after the full dataset is processed.

The e-process detects the same case sequentially during data collection, from a fraction of the data, with an empirical median stopping time of $\PoisLinearReassureEMed$ observations (Table~\ref{tab:reassurance}), corresponding to $\PoisLinearReassureEMedPct\%$ of the dataset.

\subsubsection{Effect of observation ordering}\label{sec:ordering}

As established in \S\ref{sec:method-portfolio}, the terminal statistic $E_T$ is order-invariant, so under the final-evidence rule $E_T \geq 1/\alpha$ the e-process's detection rate does not depend on the order in which observations are processed. The sequential crossing rule we use for early stopping, $\sup_{t \leq T} E_t \geq 1/\alpha$, remains type-I valid for any predictable ordering, but its stopping time and its rejection indicator can depend on the order. Empirically this dependence is small: the e-process's detection rate is nearly unchanged between random and spatial ordering, whereas the Bonferroni sequential $\chi^2$ test lacks even the terminal invariance; its $\alpha/t$ boundary tightens with $t$, so both its detection rate and stopping time depend on arrival order, and its transition-region detection rate is substantially lower under random ordering than under spatial (Appendix~\ref{sec:si-ordering}).

\subsubsection{Magnitude-accepted regime: consequences and detection}\label{sec:reassurance}

We use two settings in which Morozov accepts the fit (residual magnitudes are consistent with noise) to ask two questions: whether acceptance is costly, by measuring quantities of interest the accepted model produces, and whether the cost is avoidable, by checking which diagnostics detect the misspecification Morozov misses. We take three quantities of interest (QoIs), chosen to span the kinds of output a downstream decision typically depends on. The first is a \emph{near-boundary point prediction}, the solution at the grid node nearest $x = 0.95$, a local field value. The second is a \emph{regional average}, the mean of the nodal solution values over $[0.8, 1]$, a smoothed regional quantity. The third is the \emph{boundary flux} $q(1) = -\theta(1)\,u'(1)$ at the right boundary, a derived functional involving the gradient, evaluated with a second-order one-sided difference $u'(1) \approx (3u_N - 4u_{N-1} + u_{N-2})/(2\Delta x)$. The three aggregate the solution differently, responding differently to the same model error.

\begin{figure}
  \centering
  \includegraphics[width=0.8\textwidth]{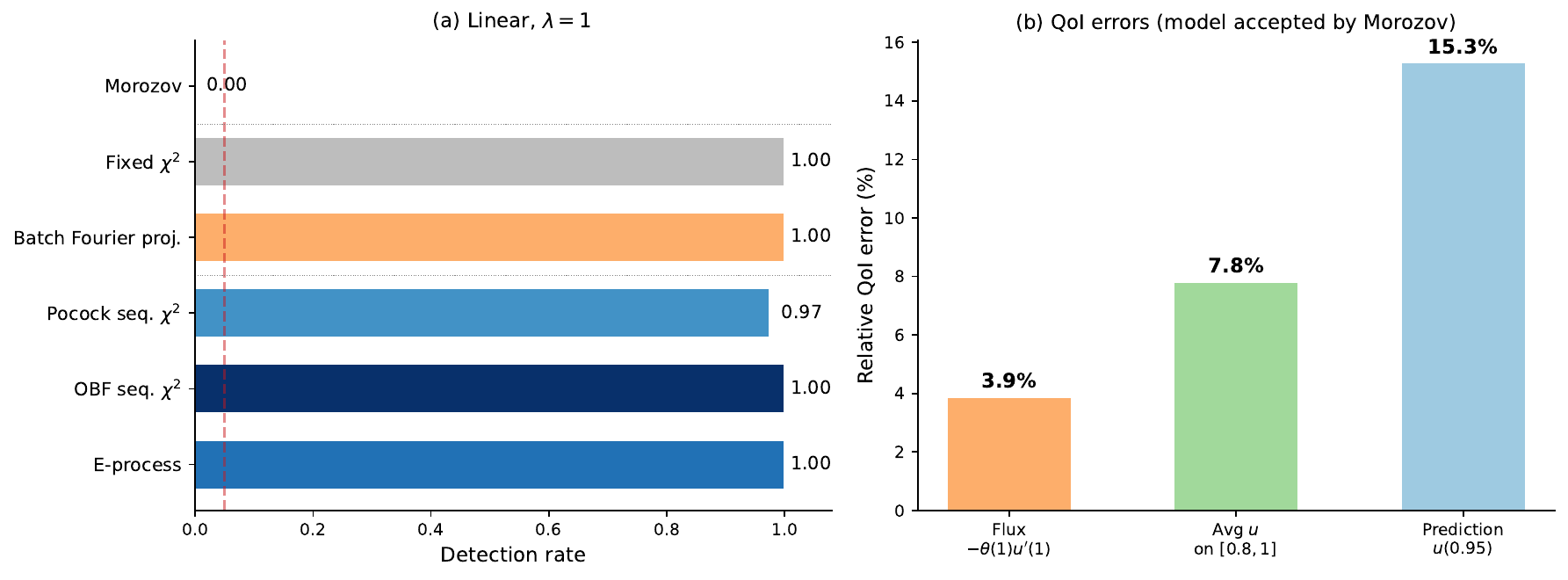}
  \caption{Diagnostic comparison for the linear case ($\lambda = 1.0$). \textbf{(a)}~Detection rates for six diagnostics arranged by category. Morozov misses the misspecification entirely; all other diagnostics achieve $\geq 0.97$ detection. The e-process and OBF do so sequentially using $\PoisLinearReassureEMed$ and $\PoisLinearReassureOBFMed$ observations respectively (Table~\ref{tab:reassurance}), compared to the full $T = 501$ observations required by the batch tests. \textbf{(b)}~QoI errors under the Morozov-accepted model.}
  \label{fig:reassurance}
\end{figure}

Table~\ref{tab:reassurance} reports two settings chosen so that residual magnitudes are consistent with noise: Morozov accepts in $100\%$ of runs, yet the accepted model's QoI errors are non-negligible and do not decrease as more data is collected (Fig.~\ref{fig:reassurance}). In the linear case ($\lambda = 1.0$) the point prediction $u(0.95)$ is wrong by $\PoisLinearReassureUerr\%$, the regional average by $\PoisLinearReassureAvgErr\%$, and the boundary flux by $\PoisLinearReassureFluxErr\%$; in the Gaussian bump case ($\lambda = 0.10$) the corresponding errors are $\PoisBumpReassureUerr\%$, $\PoisBumpReassureAvgErr\%$, and $\PoisBumpReassureFluxErr\%$. A practitioner who trusted the residual-norm check would carry whichever of these errors the application requires, undetected, into the downstream decision.
 
\begin{table}[htbp]
\centering
\caption{Detection rates and median stopping times (among detecting runs) in two settings where residual magnitudes are consistent with noise ($200$ runs, $\alpha = 0.05$, default bank). Batch tests use the full sample. Pocock and OBF are  group-sequential-inspired sequential magnitude tests (empirical type-I $\PoisHzeroPocockRate$, $\PoisHzeroOBFRate$); Bonferroni is excluded for its inflated type-I rate (Appendix~\ref{sec:si-calibration}). QoI errors are deterministic.}
\label{tab:reassurance}
\small
\begin{tabular}{@{}lcccc@{}}
\toprule
& \multicolumn{2}{c}{Bump $\lambda=0.10$} & \multicolumn{2}{c}{Linear $\lambda=1.0$} \\
\cmidrule(lr){2-3}\cmidrule(lr){4-5}
Diagnostic & Detects & Obs used & Detects & Obs used \\
\midrule
Morozov (magnitude rule) & 0.00 & --- & 0.00 & --- \\
Fixed $\chi^2$ (magnitude+batch) & 1.0 & 501 & 1.0 & 501 \\
Batch Fourier (structure+batch) & 1.0 & 501 & 1.0 & 501 \\
Pocock seq.\ $\chi^2$ (magnitude+seq) & 1.0 & \PoisBumpReassurePocockMed & \PoisLinearReassurePocockDetRate & \PoisLinearReassurePocockMed \\
OBF seq.\ $\chi^2$ (magnitude+seq) & 1.0 & \PoisBumpReassureOBFMed & 1.0 & \PoisLinearReassureOBFMed \\
E-process (structure+seq) & 1.0 & \PoisBumpReassureEMed & \PoisLinearReassureEDetRate & \PoisLinearReassureEMed \\
\midrule
QoI error: $u(0.95)$ & \multicolumn{2}{c}{\PoisBumpReassureUerr\%} & \multicolumn{2}{c}{\PoisLinearReassureUerr\%} \\
QoI error: avg $u$ on $[0.8,1]$ & \multicolumn{2}{c}{\PoisBumpReassureAvgErr\%} & \multicolumn{2}{c}{\PoisLinearReassureAvgErr\%} \\
QoI error: boundary flux $-\theta(1)u'(1)$ & \multicolumn{2}{c}{\PoisBumpReassureFluxErr\%} & \multicolumn{2}{c}{\PoisLinearReassureFluxErr\%} \\

Morozov acceptance rate & \multicolumn{2}{c}{100\%} & \multicolumn{2}{c}{100\%} \\
\bottomrule
\end{tabular}
\end{table}

\subsection{Stokes flow with unknown basal drag}\label{sec:stokes}
 
To demonstrate the same mechanism in a more ill-posed boundary-inference setting, we apply the e-process to a 2D Stokes flow problem~\citep{petra2012inexact,boulakia2013stability,rasmussen2024bayesian}, where the parameter $\beta(x)$ on the base must be inferred from velocity observations on the surface. The Stokes operator smooths information from base to surface: the base-level error in $\beta(x)$ is attenuated by the time it reaches the observation surface. This attenuation is well documented in glaciology: the base-to-surface transfer is effectively low-pass, so structurally different basal fields can produce nearly indistinguishable surface velocities~\citep{arthern2010initialization,wolovick2023regularization,martin2014adjoint}.

The Stokes equations are solved on $[0,1]\times[0,0.2]$ using Taylor--Hood finite elements on a triangular mesh obtained from a $150\times30$ rectangular subdivision, giving about $4.1\times 10^4$ degrees of freedom.
The equations are solved using DOLFINx/FEniCSx~\citep{alnaes2015fenics}. The boundary conditions are as follows: Robin condition at the base ($y = 0$, basal drag $\beta(x)\mathbf{u}$ with $\beta(x) = \exp(\sum_k \theta_k \phi_k(x))$), Neumann at the surface ($y = 0.2$, prescribed stress $\tau(x) = 10(\sin(12\pi x) + 1)$), and zero Dirichlet on lateral walls. The true parameter values are \stokesThetaTruthFormatted\ with $K = 4$ coefficients. Both surface-velocity components $(u_x,u_y)$ are observed at $N=\stokesNobs$ random surface locations with independent Gaussian noise of standard deviation $\sigma=0.01$ per component. Both components enter the least-squares fit, while the diagnostics reported below are applied only to the $u_x$ residual channel.
 
While the true basal drag uses four coefficients $(\theta_0,\ldots,\theta_3)$ in the basis of Appendix~\ref{sec:si-stokes}, we fit with only the first three $(\theta_0,\theta_1,\theta_2)$, a structural misspecification that omits the highest-index basis mode $\theta_3\phi_3$ present in the truth. The base-level error in $\beta(x)$ is attenuated by the time it reaches the observation surface, producing surface residuals with $\mathrm{RMS}/\sigma = \stokesRmsObsOverSigmaMean$, below the Morozov threshold at $\tau = 1.5$ we adopt throughout (well above the principle's nominal $\tau = 1$, and the more permissive end of the range a practitioner might choose). The dependence on $\tau$, including the insensitivity region $\|\mu^*\|/\sigma < \sqrt{\tau^2 - 1}$, is examined in \S\ref{sec:linear-subnoise}.

\begin{figure}[!htbp]
  \centering
  \includegraphics[width=0.8\textwidth]{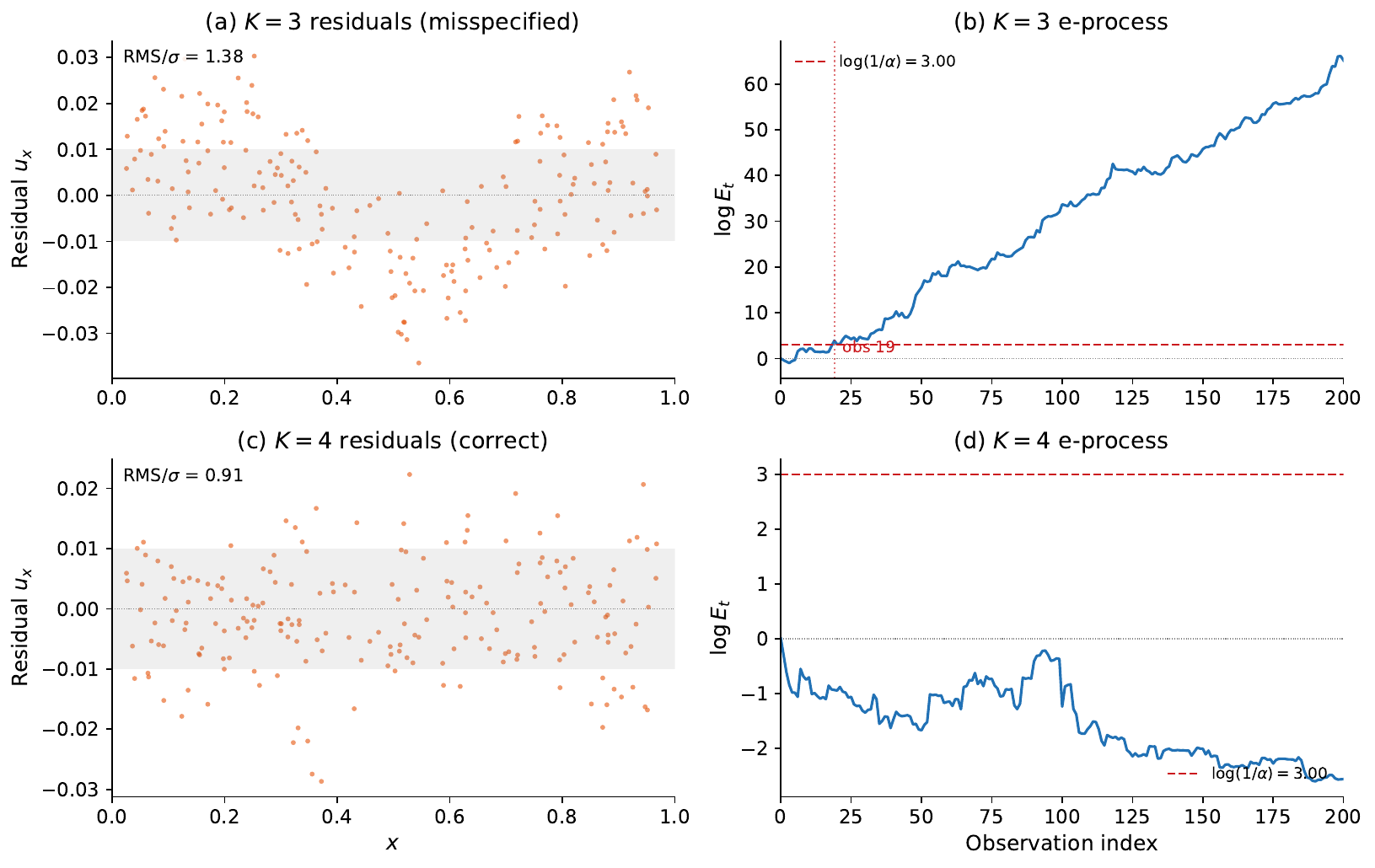}
\caption{Diagnostic failure on the 2D Stokes inverse problem; representative single runs from the $K=3$ (misspecified) and $K=4$ (correct) ensembles, near the median e-process stop time and final log-evidence respectively. \textbf{Top ($K=3$):} (a)~surface residuals, consistent with noise ($\mathrm{RMS}/\sigma \approx \stokesRmsObsOverSigmaMean$) yet visibly structured; (b)~e-process crosses the threshold early. \textbf{Bottom ($K=4$):} (c)~unstructured residuals; (d)~e-process stays below threshold throughout. Quantitative detection rates and the basal-drag consequences are given in \S\ref{sec:stokes} and Table~\ref{tab:stokes-detection}.}
  \label{fig:stokes}
\end{figure}

Fig.~\ref{fig:stokes} shows the result. The $K = 3$ residuals in panel~(a) have $\mathrm{RMS}/\sigma = \stokesRmsObsOverSigmaMean$, small enough to pass the magnitude check. The residuals do have clear spatial structure, and the e-process in panel~(b) detects it, crossing the rejection threshold; across the MC ensemble the median stopping time is \stokesEpMedianStopKThree\ observations. The correctly specified $K = 4$ model (bottom row) recovers the true basal drag, produces unstructured residuals, and the e-process correctly does not reject.

The diagnostic outcome is summarised in Table~\ref{tab:stokes-detection}. Morozov accepts the misspecified fit in nearly every run; the batch tests detect only after the full record; and all three sequential tests detect from a fraction of the surface observations, the e-process earliest (median \stokesEpMedianStopKThree\ of \stokesNobs, \stokesEpMedianStopFractionKThree\ of the data) and at a nominal null rate, whereas OBF reaches comparable power only with a mildly inflated null (\stokesNullOBF). 

\begin{table}[htbp]
\centering
\caption{Diagnostic comparison on the 2D Stokes basal-drag inverse problem (\stokesNmcKThree\ Monte Carlo runs at $K = 3$; \stokesNmcKFour\ runs for the $K = 4$ null). Surface velocity is observed at $N = \stokesNobs$ locations. ``Median stop'' is the median stopping time among detecting runs, reported as observations and as a fraction of $N$; batch tests use the full record. The null column is the empirical type-I rate under the correctly specified $K = 4$ fit ($\alpha = 0.05$).}
\label{tab:stokes-detection}
\small
\begin{tabular}{@{}lccc@{}}
\toprule
Diagnostic & Power ($K=3$) & Median stop & Null type-I ($K=4$) \\
\midrule
Morozov (magnitude rule)                & \stokesMorozovDetKThree      & ---                       & \stokesNullMorozov \\
Fixed-sample $\chi^2$ (magnitude+batch) & \stokesFixedChiSqDetKThree   & \stokesNobs/\stokesNobs   & --- \\
Batch Fourier (structure+batch)         & \stokesBatchFourierDetKThree & \stokesNobs/\stokesNobs   & \stokesNullBatchFourier \\
Pocock seq.\ $\chi^2$ (magnitude+seq)   & \stokesPocockDetKThree       & \stokesPocockMedianStopKThree/\stokesNobs & \stokesNullPocock \\
OBF seq.\ $\chi^2$ (magnitude+seq)      & \stokesOBFDetKThree          & \stokesOBFMedianStopKThree/\stokesNobs    & \stokesNullOBF \\
E-process (structure+seq)               & \stokesEpDetKThree           & \stokesEpMedianStopKThree/\stokesNobs\ (\stokesEpMedianStopFractionKThree) & \stokesNullEp \\
\bottomrule
\end{tabular}
\end{table}

Note that since the forward operator is smoothing, a structurally wrong $\beta(x)$ is nearly indistinguishable from the truth in surface velocity, yet the inferred field is qualitatively wrong: the location of minimum basal drag, the quantity most relevant to grounding-line stability, is displaced from $x = \stokesBetaArgminTruth$ to $x = \stokesBetaArgminKThree$. This is not an artefact of a poor or under-regularised fit. The $K = 3$ parameters are fit optimally to the noise-free truth (\S\ref{sec:robustness}), and the failure is structural: a regulariser controls the amplitude of $\beta$ but cannot restore a harmonic the model class omits, nor move the drag minimum to where the surface data cannot resolve it. The e-process flags the inadequacy, before the full dataset is observed.

\subsection{Ice-stream basal drag inversion using \textit{icepack}}\label{sec:icepack}

To verify that the failure mode persists in a more realistic forward operator from a community modelling package, we apply the e-process to a basal-friction inversion implemented in \textit{icepack}~\citep{shapero2021icepack}, a community finite-element ice-flow model built using the Firedrake FEM library~\citep{rathgeber2016firedrake}. This is a controlled synthetic benchmark: the true friction is a known $K = 4$ coefficient profile, the practitioner-chosen model class is $K = 3$, observations are simulated by adding noise to the forward solve. 
The setup uses the shallow shelf approximation (SSA) on a fully grounded rectangular ice stream domain $[0, L_x] \times [0, L_y]$ with $L_x = 50$~km, $L_y = 12$~km, and bed and surface elevations sloping linearly along the flow direction (full geometry in Appendix~\ref{sec:si-icepack}). The basal friction parameterisation uses a related truncated-Fourier construction, with a different frequency indexing from the Stokes case: $C(x) = C_0\exp(\theta_0 + \sum_{k=1}^{K-1}\theta_k\phi_k(x/L_x))$. Surface velocity is observed at $N = 200$ random locations in the interior with noise $\sigma = 1.3$~m/yr (approximately $1.4\%$ of the mean speed, comparable to the precision of high-quality satellite ice-velocity products in slow-flow regions~\citep{recinos2025mapping}).

The forward operator differs from the Stokes case: rather than smoothing vertically from base to surface, the SSA system smooths horizontally through the membrane stress term $\nabla\cdot(hM)$, attenuating high-wavenumber components of $C(x)$ in the depth-averaged velocity field. The mechanism is the same: a structured error in the basal coefficient is hidden in the surface observations. The expert bank and rejection threshold are identical to the Poisson and Stokes cases.
\begin{figure}
  \centering
  \begin{subfigure}[ht]{0.32\textwidth}
    \centering
    \includegraphics[width=0.8\textwidth]{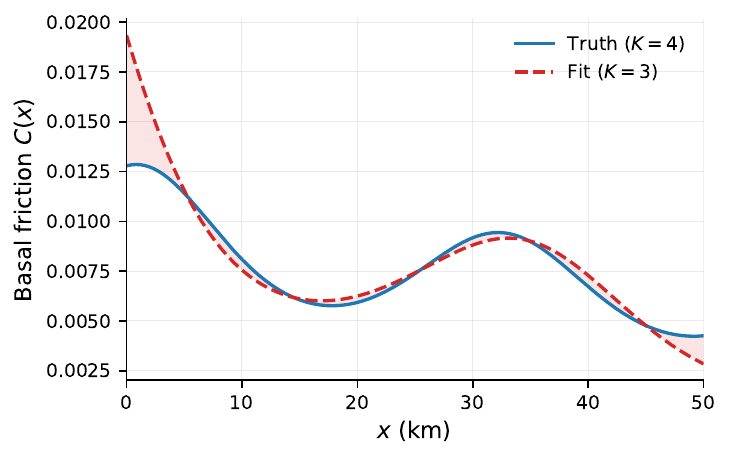}
    \caption{Basal friction $C(x)$: $K = 3$ fit (dashed) vs.\ truth (solid). Pointwise error reaches $\IceBasalFrictionErrMaxPct\%$, concentrated near the inflow boundary.}
    \label{fig:icepack:a}
  \end{subfigure}
  \hfill
  \begin{subfigure}[t]{0.32\textwidth}
    \centering
    \includegraphics[width=0.8\textwidth]{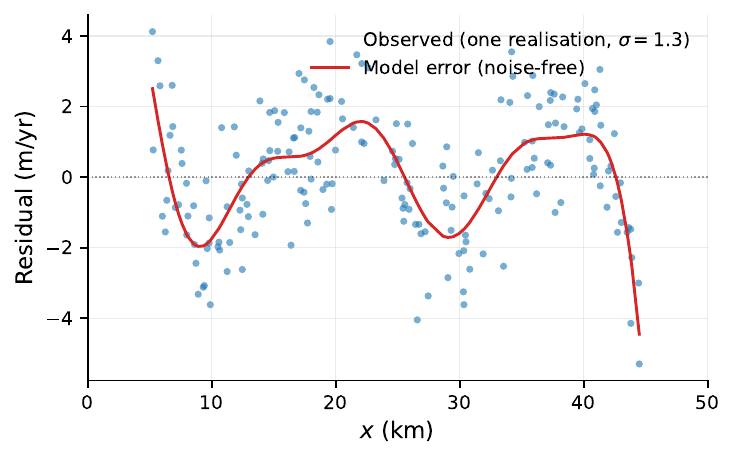}
    \caption{Surface velocity residuals with $\sigma = 1.3$~m/yr. The noise-free model error (red) is comparable in magnitude to the noise but spatially structured; $\mathrm{RMS}/\sigma = \IceERMS$.}
    \label{fig:icepack:b}
  \end{subfigure}
  \hfill
  \begin{subfigure}[t]{0.32\textwidth}
    \centering
    \includegraphics[width=0.8\textwidth]{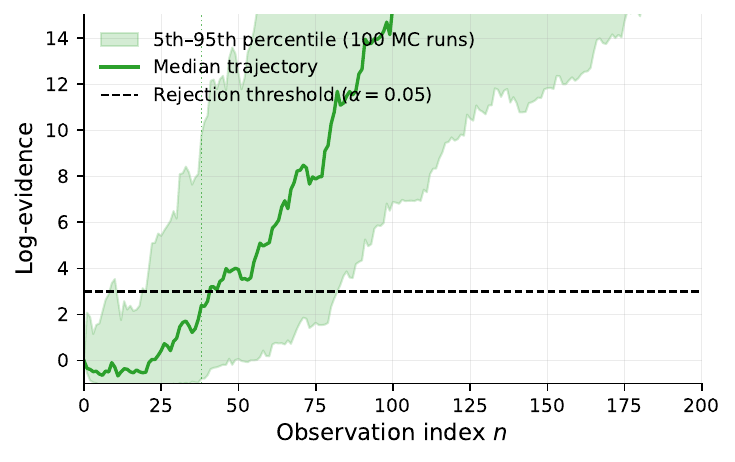}
    \caption{E-process trajectories over 100 noisy-refit Monte Carlo runs: median log-evidence (line) and the $5$th--$95$th percentile band. The median stopping time among detecting runs is observation~$\IceEMed$ of 200.}
    \label{fig:icepack:c}
  \end{subfigure}
  \caption{Diagnostic failure on a glaciological SSA inverse problem (\textit{icepack}). The horizontal forward-operator smoothing decouples observation-space fit from parameter-space accuracy: surface velocities match truth to median $\IceSurfaceVelErrMedianPct\%$ in the interior, but the inferred basal friction is wrong, and the predicted terminus ice flux is $\IceTerminusFluxErrPct\%$ too high. Null type-I rates (residuals from a correctly specified $K = 4$ fit) are well below nominal $\alpha = 0.05$: e-process $\IceENullRate$, batch Fourier $\IceBFNullRate$.}
  \label{fig:icepack}
\end{figure}

Fig.~\ref{fig:icepack} shows the result. The $K = 3$ fit reproduces surface velocity to within median $\IceSurfaceVelErrMedianPct\%$ of truth in the interior (panel~b shows residuals of order $\sigma$), passing Morozov's check at $\mathrm{RMS}/\sigma = \IceERMS$. Yet the inferred basal friction (panel~a) deviates from truth by up to $\IceBasalFrictionErrMaxPct\%$ pointwise, with the largest discrepancy at the inflow boundary. Across the Monte Carlo ensemble, the e-process has median stopping time $\IceEMed$, detecting the structural inadequacy from approximately a fifth of the data.

The detection comparison is given in Table~\ref{tab:icepack}. As in the Stokes case, Morozov accepts the misspecified fit while every other diagnostic detects it, the batch tests only after the full record and the sequential tests from a fraction of it, all reported null rates are at or below the nominal level. Bonferroni is omitted because its null rejection rate is \IceBonfNullRate. Unlike the subnoise Poisson and Stokes regimes, however, the e-process holds no speed advantage here: the icepack residuals are super-noise (post-noise $\mathrm{RMS}/\sigma = \IceERMS$), so the magnitude-based sequential tests have ample per-observation signal, and Pocock stops earlier than the e-process (medians \IcePocockMed\ and \IceEMed, respectively). The value of the e-process on this problem is therefore not earlier detection but its wealth-based attribution and the anytime-valid monitoring guarantee available when the fitted model is fixed independently of the monitored residuals. For the same-data \textit{icepack} pipeline evaluated here, calibration is empirical.

\begin{table}[htbp]
\centering
\caption{Diagnostic comparison on the \textit{icepack} ice-stream inverse problem ($100$ runs at $K=3$; $100$ for the $K=4$ null, $\alpha = 0.05$). ``Median stop'' is observations\,/\,$N$ among detecting runs; batch tests use the full record. Bonferroni is excluded for its inflated null rate $\IceBonfNullRate$ (Appendix~\ref{sec:si-icepack}). The QoI is terminus ice flux predicted by the accepted $K=3$ model.}
\label{tab:icepack}
\small
\begin{tabular}{@{}lccc@{}}
\toprule
Diagnostic & Power & Median stop & Null type-I \\
\midrule
Morozov                       & \IceMrzvDet\%   & ---            & \IceMrzvNullRate \\
Fixed-sample $\chi^2$         & \IceFCDet\%     & 200/200        & --- \\
Batch Fourier                 & \IceBFDet\%     & 200/200        & \IceBFNullRate \\
Pocock seq.\ $\chi^2$         & \IcePocockDet\% & \IcePocockMed/200 & \IcePocockNullRate \\
OBF seq.\ $\chi^2$            & \IceOBFDet\%    & \IceOBFMed/200    & \IceOBFNullRate \\
E-process                     & \IceEDet\%      & \IceEMed/200      & \IceENullRate \\
\midrule
Surface velocity error (interior) & \multicolumn{3}{c}{median $\IceSurfaceVelErrMedianPct\%$} \\
Pointwise basal friction error    & \multicolumn{3}{c}{up to $\IceBasalFrictionErrMaxPct\%$} \\
Terminus ice flux QoI error       & \multicolumn{3}{c}{$\IceTerminusFluxErrPct\%$} \\
\bottomrule
\end{tabular}
\end{table}

\section{From testing to model repair}\label{sec:correction}
 
The e-process provides more than a binary reject/accept decision. After rejection, the portfolio wealth distribution across experts attributes the evidence to dominant spatial residual patterns in the expert bank. This diagnostic is the central output of the pipeline: it suggests which spatial modes are driving the misspecification before any correction is attempted. We then show that fitting a parsimonious correction on these patterns recovers the practitioner's quantity of interest at performance comparable to a brute-force regression on the full basis, with the caveat that very small expert sets (\emph{e.g.} top 3) underfit and can worsen the error on the QoI. We frame the correction step as exploratory: detection carries a formal type-I error guarantee under the fixed-fit monitoring conditions, but the subsequent diagnosis and correction are practical heuristics whose validity as a follow-on adequacy check would require held-out data or an independent e-process.
 
We also draw connections to the model discrepancy literature. \citet{kennedy2001bayesian} model reality as $y(x) = G_\theta(x) + \delta(x) + \epsilon$, where $\delta(x) \sim \mathrm{GP}(0, k_\ell)$ is a Gaussian process discrepancy on the model output, and \citet{brynjarsdottir2014learning} show that omitting $\delta$ biases the inferred $\theta$. In that framework $\delta$ is \emph{assumed to exist} and a full posterior over it is inferred jointly with $\theta$, typically by MCMC with the forward model in the loop. Our experts are a discrete basis for this output discrepancy: each proposes $\delta(x) = a\,\phi_k(x)$, and the universal portfolio integrates over these proposals, analogously to the GP prior on $\delta$; however, we do not infer a posterior over $\delta$. We first \emph{test} whether $\delta = 0$ (detection), then read off the dominant patterns from the portfolio (diagnosis), and only then fit a deterministic point correction (repair). 
 
The correction is computed retrospectively, after all observations have been processed: we rank the experts by their cumulative log-likelihood ratio $L_{T,k}$ over the full trajectory and collapse the signed-amplitude copies, keeping the highest-scoring variant of each unique shape, to obtain a ranked list of $13$ shapes. The expert bank carries each spatial shape at twelve signed amplitudes (\S\ref{sec:experts}); this amplitude grid is what lets the test detect an error of unknown magnitude and either sign, but it is redundant for repair, where the amplitude of each shape is re-estimated directly. Taking the top $K_c$, we fit the discrepancy $\hat\delta(x) = \sum_{k \in \mathrm{top}} \beta_k \phi_k(x)$ by ridge regression on the full residual record ($\lambda_{\mathrm{ridge}} = 10^{-6}$).

We evaluate this detect--diagnose--correct pipeline on the three Poisson misspecifications for which the accepted model has materially wrong quantities of interest (linear $\lambda = 1.0$, bump $\lambda = 0.1$, three-step $\lambda = 0.15$).
We report the two QoIs, $u(0.95)$ and the regional average on $[0.8,1]$, both functionals of $u$; we omit the boundary flux, which depends on the diffusivity $\theta$ that a solution-space correction leaves unchanged.
Table~\ref{tab:correction} compares the uncorrected null, e-informed corrections using the top $K_c \in \{3, 5, 10\}$ shapes, and a blind regression on all $13$ shapes, against the oracle (the true discrepancy). The e-informed correction with $10$ shapes reduces the $u(0.95)$ error from $\CorrBumpNoCorrUerr$--$\CorrThreeStepNoCorrUerr\%$ uncorrected to $\CorrLinearInfTenUerr$--$\CorrBumpInfTenUerr\%$, matching the blind full-basis regression, and brings the regional-average error to roughly $1\%$; with only $5$ shapes it achieves  $\CorrLinearInfFiveUerr$--$\CorrBumpInfFiveUerr\%$ on $u(0.95)$. The top-3 correction is less reliable. It improves both QoIs for the linear and three-step cases, but for the bump it increases the $u(0.95)$ error from $\CorrBumpNoCorrUerr\%$ to $\CorrBumpInfThreeUerr\%$ and the regional-average error from $\CorrBumpNoCorrAvgErr\%$ to $\CorrBumpInfThreeAvgErr\%$. This bump-case failure is consistent with underfitting: with only three selected shapes, the regression attributes residual signal to an incomplete basis and displaces the QoI further from truth. Using the top five or ten shapes avoids this failure in these experiments, but the correction step remains a heuristic rather than a calibrated procedure.

\begin{table}[htbp]
\centering
\caption{Detect--Diagnose--Correct: mean relative QoI error across correction strategies (200 runs each, oracle fit, default bank $K = 156$). E-informed corrections use the top-$k$ unique spatial shapes ranked by cumulative log-likelihood ratio at the end of the trajectory.}
\label{tab:correction}
\small
\begin{tabular}{@{}lcccccc@{}}
\toprule
& \multicolumn{2}{c}{Linear $\lambda = 1$} & \multicolumn{2}{c}{Bump $\lambda = 0.1$} & \multicolumn{2}{c}{Three-step $\lambda = 0.15$} \\
\cmidrule(lr){2-3}\cmidrule(lr){4-5}\cmidrule(lr){6-7}
Strategy & $u(.95)$ & avg & $u(.95)$ & avg & $u(.95)$ & avg \\
\midrule
No correction & 15.3\% & 7.8\% & 12.0\% & 5.4\% & 20.1\% & 10.7\% \\
E-informed (3) & 7.2\% & 1.2\% & 16.9\% & 8.0\% & 10.2\% & 3.5\% \\
E-informed (5) & 3.4\% & 1.4\% & 4.6\% & 1.6\% & 4.0\% & 1.7\% \\
E-informed (10) & 3.8\% & 1.1\% & 4.5\% & 1.2\% & 4.2\% & 1.1\% \\
Blind (all) & 4.2\% & 1.1\% & 4.8\% & 1.2\% & 5.2\% & 1.1\% \\
Oracle & 0.0\% & 0.0\% & 0.0\% & 0.0\% & 0.0\% & 0.0\% \\
\bottomrule
\end{tabular}
\end{table}

\begin{figure}
  \centering
  \includegraphics[width=0.8\textwidth]{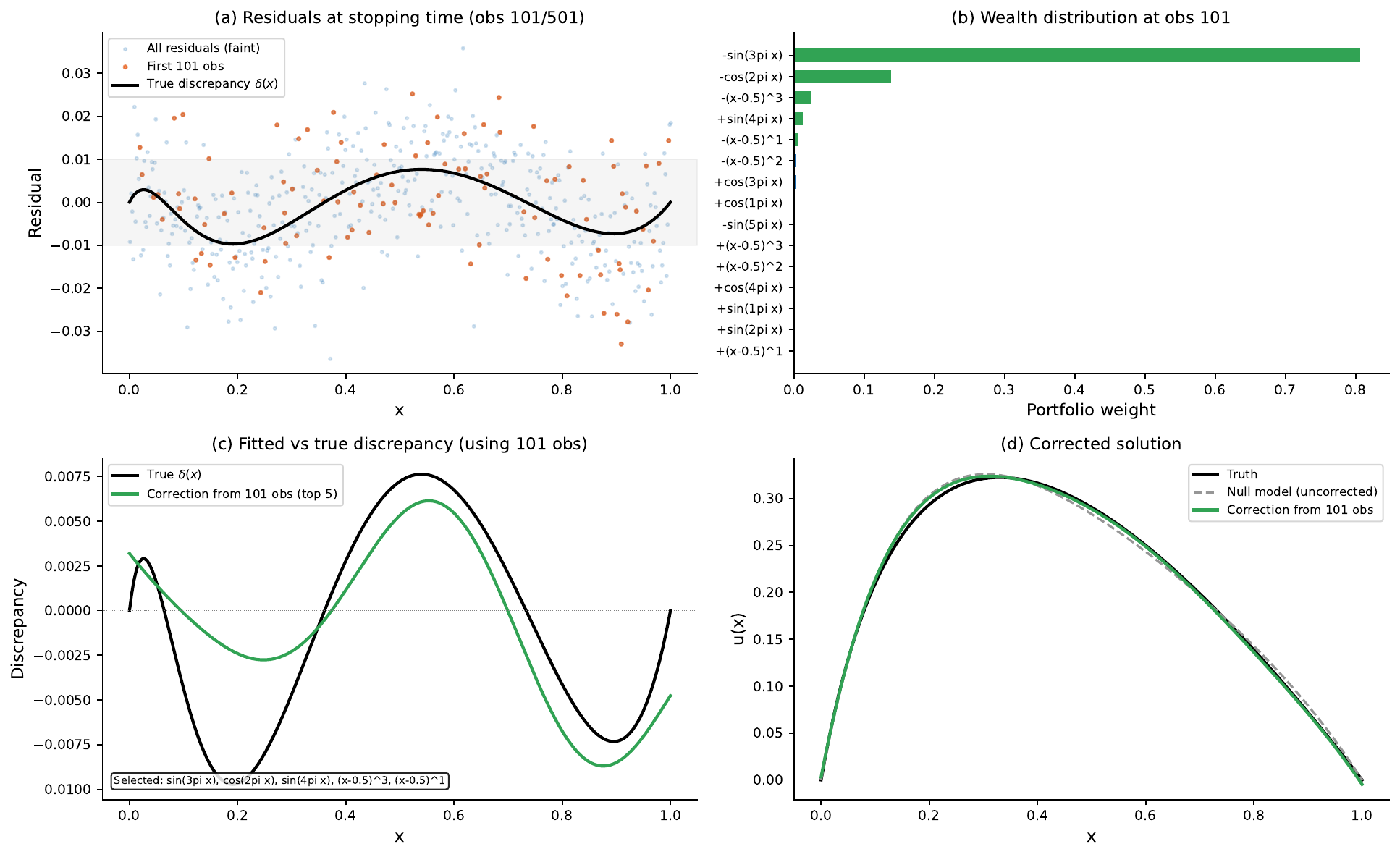}
  \caption{Detect--Diagnose--Correct on the linear case ($\lambda = 1.0$), single run at the stopping time ($101$ of $501$). \textbf{(a)}~Residuals available at rejection (orange) against the true discrepancy $\delta(x)$ (black), within the noise band. \textbf{(b)}~Portfolio wealth concentration. \textbf{(c)}~Discrepancy fitted from the top-$5$ shapes vs.\ truth. \textbf{(d)}~The corrected solution recovers the QoI region $[0.8,1]$.}
  \label{fig:pipeline}
\end{figure}

Fig.~\ref{fig:pipeline} illustrates a single run on the linear case in the Poisson problem: at the stopping time the wealth concentrates on two experts (Fig.~\ref{fig:pipeline} b). Across runs the ranking is stable: the three most frequently selected unsigned shapes are $\sin(3\pi x)$ (\PoisLinearTopThreeFirstCount/\PoisLinearTopThreeNRuns\ runs), $\cos(2\pi x)$ (\PoisLinearTopThreeSecondCount/\PoisLinearTopThreeNRuns), and $\cos(4\pi x)$ (\PoisLinearTopThreeThirdCount/\PoisLinearTopThreeNRuns); the same modes dominate when the correction is fit on different random subsets of the data. The portfolio wealth therefore reliably attributes the evidence to specific residual modes in the expert bank, providing interpretable guidance for subsequent model refinement.
 
\section{Robustness studies}\label{sec:robustness}
\subsection{Oracle vs.\ noisy-data fit}
The best-fit null parameter $\hat\theta$ is computed by minimising the squared error against the true (noiseless) misspecified solution, an ``oracle'' procedure that gives the null model every advantage. In practice, $\hat\theta$ would be estimated from noisy observations. Table~\ref{tab:oracle} compares the two approaches for both the Poisson and Stokes problems.
Results are essentially identical: e-process median stopping times change by fewer than $10$ observations across the three Poisson settings, and detection rates change by at most one percentage point. This robustness arises because least-squares estimation with hundreds of observations averages out the noise, yielding effectively the same $\hat\theta$ in both cases. These comparisons support the numerical stability of the diagnostic under noisy refitting, but the formal anytime-valid guarantee stated in Section~\ref{sec:methods} remains the conditional fixed-\(\hat\theta\) guarantee unless independent monitoring data, sample splitting, or universal inference is used.

\begin{table}[htbp]
\centering
\caption{Oracle vs.\ noisy-data fit comparison ($\alpha = 0.05$, default bank $K = 156$). Poisson cells use 200 runs each; Stokes uses 100 oracle runs and 25 noisy-fit runs. Median stop is the median over detecting runs. Stokes noisy-fit MC is capped at $N_\mathrm{val} = 25$ replications (versus $N = 100$ for oracle) because each refit triggers a full Nelder-Mead optimisation over the Stokes forward operator.}
\label{tab:oracle}
\small
\begin{tabular}{@{}llcc@{}}
\toprule
Setting & Fit & E-proc det. & Median stop \\
\midrule
\multirow{2}{*}{Poisson linear $\lambda = 1$} & Oracle & 1.000 & 44.5 \\
& Noisy & 1.000 & 52 \\
\addlinespace
\multirow{2}{*}{Poisson bump $\lambda = 0.1$} & Oracle & 1.000 & 23 \\
& Noisy & 1.000 & 21 \\
\addlinespace
\multirow{2}{*}{Poisson three-step $\lambda = 0.15$} & Oracle & 1.000 & 18 \\
& Noisy & 1.000 & 17.5 \\

\addlinespace

Stokes $K = 3$ & Oracle & 1.0 & 19 \\
 & Noisy & 1.0 & 18 \\
\bottomrule
\end{tabular}

\end{table}

\subsection{Noise-level sensitivity}
The e-process requires a known noise level $\sigma$; Table~\ref{tab:sigma} reports its sensitivity when the assumed $\sigma_a$ differs from the true $\sigma_t$. The direction of the asymmetry is structural. Each increment~\eqref{eq:lr} is calibrated for variance $\sigma_a^2$, so under the true null its exponential has expectation $\exp\!\big(\tfrac{\mu_k^2}{2\sigma_a^4}(\sigma_t^2 - \sigma_a^2)\big)$. Overestimating $\sigma$ ($\sigma_a > \sigma_t$) makes this factor less than one: each $\exp(L_{t,k})$, and hence the mixture, remains a nonnegative supermartingale, so Ville's inequality still bounds the type-I error at $\alpha$; the test stays valid, and is conservative, as is generally the case for e-value tests~\citep{ramdas2025hypothesis}. The cost for overestimation is power, since the negative drift is steeper than calibrated. Underestimating $\sigma$ reverses the sign: the factor exceeds one, the supermartingale property is lost, and the type-I guarantee no longer holds. If the expert bank is held fixed and $\sigma_a=\sigma_t/k$, so that the assumed noise level is too small by a factor $k>1$, then the denominator in each likelihood-ratio increment is smaller by a factor $k^2$, and the increment is multiplied by $k^2$. When the amplitude grid is itself rescaled with $\sigma_a$, the same qualitative inflation occurs, but the increments are no longer a pure $k^2$ rescaling.

In this experiment, overestimating $\sigma$ does not reduce the
detection rate over the tested range: power remains $1.000$ even at
$\sigma_a=2\sigma_t$. Its observed cost is delayed detection, with the
median stop increasing from $49$ at the correctly specified noise level
to $247.5$ at twice the true noise level. Underestimating $\sigma$
instead inflates the false-alarm rate sharply.
In practice, when $\sigma$ is uncertain, use a conservative upper-bound estimate. A modest overestimate primarily costs detection speed in the tested range while keeping type-I control exact, whereas a comparable underestimate forfeits the guarantee.


\begin{table}[htbp]
\centering
\caption{Sensitivity to assumed noise level (200 runs each for $H_0$ and $H_1$: linear $\lambda = 1.0$). True $\sigma = 0.01$. Stopping time is median over detecting runs.}
\label{tab:sigma}
\small
\begin{tabular}{@{}ccccc@{}}
\toprule
$\sigma_{\mathrm{assumed}}/\sigma_{\mathrm{true}}$ & $\sigma_{\mathrm{assumed}}$ & Type-I & Power & Median stop \\
\midrule
0.50 & 0.0050 & 0.945 & 1.000 & 4.5 \\
0.80 & 0.0080 & 0.235 & 1.000 & 26.0 \\
0.90 & 0.0090 & 0.060 & 1.000 & 36.0 \\
1.00 & 0.0100 & 0.010 & 1.000 & 49.0 \\
1.10 & 0.0110 & 0.000 & 1.000 & 69.5 \\
1.20 & 0.0120 & 0.000 & 1.000 & 85.5 \\
1.50 & 0.0150 & 0.000 & 1.000 & 137.0 \\
2.00 & 0.0200 & 0.000 & 1.000 & 247.5 \\
\bottomrule
\end{tabular}
\end{table}

\subsection{Expert bank sensitivity}\label{sec:bank-sensitivity}

The default bank uses 5 Fourier frequencies and 3 polynomial degrees at 12 signed amplitudes for $K = 156$ experts. The mixture inequality bounds the regret of including extra experts at $\log K / t$ per observation, predicting near-insensitivity to bank size above a problem-dependent threshold and severe power loss below it. Table~\ref{tab:bank} examines these predictions on three cases at $\alpha = 0.05$: the Poisson linear $\lambda = 1$ case (\S\ref{sec:poisson}), the Poisson bump $\lambda = 0.1$ case (\S\ref{sec:poisson}), and the icepack $K = 3$ fit (\S\ref{sec:icepack}). Four bank configurations are compared: a small bank (3 Fourier frequencies, 2 polynomial degrees, $K = 96$), the default bank ($K = 156$), a large bank (8 frequencies, 4 polynomial degrees, $K = 240$), and a Fourier-only bank that retains the default frequencies but excludes all polynomial experts ($K = 120$).


\begin{table}[htbp]
\centering
\caption{Expert bank sensitivity across PDE problems (200 Monte Carlo runs for Poisson problems, 100 runs for \textit{icepack}, $\alpha = 0.05$). Median stopping times among detecting runs.}
\label{tab:bank}
\small
\begin{tabular}{@{}llcccc@{}}
\toprule
Case & Bank & Shapes & $K$ & Median stop & EP det \\
\midrule
\multirow{4}{*}{Poisson linear $\lambda = 1$} & Small (3 freq, 2 poly) &  8 &  96 &   80.0 & 100\% \\
 & Default (5 freq, 3 poly) & 13 & 156 &   80.0 & 100\% \\
 & Large (8 freq, 4 poly) & 20 & 240 &   77.0 & 100\% \\
 & Fourier only (5 freq) & 10 & 120 &   80.0 & 100\% \\
\midrule
\multirow{4}{*}{Poisson bump $\lambda = 0.1$} & Small (3 freq, 2 poly) &  8 &  96 &   20.0 & 100\% \\
 & Default (5 freq, 3 poly) & 13 & 156 &   20.0 & 100\% \\
 & Large (8 freq, 4 poly) & 20 & 240 &   20.0 & 100\% \\
 & Fourier only (5 freq) & 10 & 120 &   20.0 & 100\% \\
\midrule
\multirow{4}{*}{icepack $K = 3$} & Small (3 freq, 2 poly) &  8 &  96 &    --- & 25\% \\
 & Default (5 freq, 3 poly) & 13 & 156 &   37.0 & 100\% \\
 & Large (8 freq, 4 poly) & 20 & 240 &   26.0 & 100\% \\
 & Fourier only (5 freq) & 10 & 120 &   33.0 & 100\% \\
\bottomrule
\end{tabular}
\end{table}

Two regimes are visible. For both Poisson cases, all four banks achieve $100\%$ detection with similar stopping times. On \textit{icepack}, the default, large, and Fourier-only banks achieve $100\%$ detection, with median stops $37$, $26$, and $33$, respectively. The small bank, which includes only Fourier frequencies $j=1,2,3$ and therefore omits the dominant fifth-frequency residual mode, detects in only $25\%$ of runs. Thus bank size itself is not decisive; what matters is coverage of the
dominant residual mode.

Examining the leading components of the wealth distribution averaged over Monte Carlo runs explains the asymmetry. On the Poisson bump case, both the default and Fourier-only banks place approximately $99\%$ of their mean wealth on the signed $-\sin(3\pi x)$ pattern, with effective bank size about $1.1$. Both have median stopping time $20$, showing that polynomial experts are not needed for this case under the signed bank.

On \textit{icepack} the wealth distribution is sharper still. Both the default and Fourier-only banks place essentially all their wealth on a single shape, $\cos(5\pi x/L_x)$, with effective bank size $\approx 1.0$ and stopping times within Monte Carlo error of each other. Polynomial experts receive no appreciable weight on this problem. Empirically, the residuals of the misspecified $K=3$ fit have their strongest signature at $\cos(5\pi x/L_x)$, even though the coefficient omitted from the friction parameterisation is the $\cos(3\pi x/L_x)$ mode; the lower retained coefficients appear to absorb part of the omitted structure during refitting, leaving a higher-frequency residual component (\S\ref{sec:si-icepack}). The small bank, whose maximum Fourier frequency is $\cos(3\pi x/L_x)$, has no expert at this residual mode, so its detection power is much lower.

The practical implication is sharper than a uniform robustness claim. The bank must include shape classes that cover the lowest-order modes by which the misspecification differs from the null, where ``lowest-order'' is set by the geometry of the problem rather than by a universal heuristic. In practice this means including Fourier modes up to a frequency that resolves the spatial scale of expected residual structure (a domain-knowledge input, not a tuning parameter) and low-degree polynomials to capture trend. The wealth distribution provides direct diagnostic feedback when the bank is misaligned; \emph{e.g.} the sharply reduced $25\%$ detection rate of the small icepack bank signals an inadequate basis. A default bank that errs on the side of inclusion, as our $K = 156$ does, avoids the failure mode at negligible cost.
 
\section{Discussion and conclusions}\label{sec:discussion}

This work proposes a structure-sensitive sequential diagnostic for PDE inverse problems. 
Two features make the e-process straightforward to adopt. The algorithm is a post-processing step on the residuals of an existing fit: it requires no change to the forward solver, no refitting, and the per-observation update is $O(K)$ arithmetic, negligible against a single PDE solve. Since the wealth is carried by named experts, the output is directly interpretable: the ranked wealth distribution indicates which spatial patterns are prominent in the residual. The expert bank is the one problem-specific choice, but for PDE inverse problems it is a natural one. The spatial domain carries a canonical inner product, the smoothness of the forward operator suggests which modes the data can resolve, and a smooth low-frequency basis of Fourier and polynomial shapes spans the discrepancies encountered in our experiments. The bank must cover the lowest-order modes by which the misspecification can differ from the null (\S\ref{sec:bank-sensitivity}), but the cost of erring toward inclusion is logarithmic in the number of experts, so a generous default suffices in these experiments.

Unlike the magnitude-based and batch comparators, the e-process provides structure sensitivity, anytime validity under optional continuation, and wealth-based attribution to spatial residual modes. From a single pass through the data it supports three steps: detection, with conditional type-I error control before the full dataset is collected; diagnosis, in which the portfolio wealth distribution attributes the evidence to spatial residual patterns; and correction, in which the top-weighted patterns suggest a cheap discrepancy fit. Under the fixed-fit monitoring conditions of \S\ref{sec:methods}, detection carries a formal type-I guarantee; the diagnosis and correction are exploratory by-products of the same construction, available at negligible cost. The detection rate is competitive across all our experiments and it performs particularly well in the smooth subnoise linear regime: there the model error is far below the noise but concentrated in a few modes, magnitude tests have negligible per-observation noncentrality, and the e-process detects as reliably as any sequential test and faster, while retaining the wealth-based attribution the magnitude-based and batch comparators do not provide.

\paragraph{Posterior predictive checks.} The canonical Bayesian goodness-of-fit tool is the posterior predictive check (PPC)~\citep{gelman1996posterior}: fit a posterior over $\theta$, draw replicate datasets, and compare a test discrepancy on the replicates against the observed data. With a structure-sensitive discrepancy (for instance, the maximum projection coefficient on a Fourier basis), PPC should detect the same misspecifications the e-process detects; with a magnitude-based discrepancy, PPC inherits the insensitivity of the fixed-sample $\chi^2$ test. A full PPC comparison would require specifying a posterior workflow over $\theta$ (typically MCMC) and running many additional forward solves, which is outside our post-fit, point-estimate diagnostic scope: the e-process is designed for the practical regime in which only a single fitted $\hat\theta$ is available, and its distinctive contributions (sequential detection during data acquisition, the wealth-based attribution diagnostic, and negligible computational cost on top of the inversion itself) are visible in this regime.

\subsection{Limitations} \label{sec:limitations}

The e-process treats $\hat\theta$ as fixed once estimated; the anytime-valid type-I guarantee is therefore for residuals from a given parameter fit and for analyst-chosen stopping rules. This scope covers our two emphasised use cases (early detection from a fitted model during ongoing data collection, and continuous monitoring of a deployed model). It does not cover online refitting: a procedure in which $\hat\theta_t$ is updated from incoming data while preserving the supermartingale property unconditionally requires the universal inference framework~\citep{wasserman2020universal} and is left as future work.

The framework does not require the discrepancy to be stationary. Detection needs only that some expert has positive expected log-likelihood growth against the model error (as in \S\ref{sec:method-portfolio}). The bump and piecewise cases are spatially localized rather than translation-invariant, yet are still detected, because the Fourier--polynomial bank contains low-frequency shapes with positive alignment to their residual means. The bank is a convenient default for the smooth discrepancies of our examples, not a requirement: for sharply non-stationary or non-smooth discrepancies the natural basis differs (a wavelet or localized basis), and the construction is unchanged under that substitution. Two cases fall outside the present scope. Firstly, heteroscedastic noise breaks the constant-$\sigma$ assumption of the increment~\eqref{eq:lr}, and would require a spatially varying normalisation. Secondly, a discrepancy that drifts as observations arrive is detected but not cleanly attributed, since the dominant expert may change along the trajectory; tracking such a discrepancy connects to the online-refitting extension left to future work.

Our experiments use unregularised least-squares fits with low-dimensional parameter spaces (2--4 parameters). For regularised inversion (Tikhonov, Bayesian MAP, sparse inverse problems), the regularisation bias shifts $\hat\theta$ away from the least-squares estimate and induces residual structure of its own. The diagnostic is still meaningful in this setting, but the null hypothesis must be adjusted: ``residuals are consistent with the noise model and the expected regularisation bias'' rather than ``residuals are pure noise.'' Building this adjusted null connects to the Ingster--Marteau theory of testing under specific regularisation schemes~\citep{ingster2014signal,marteau2015unified,kroll2019rate} and is a natural direction for extending the framework to regularised inverse problems.

\bibliographystyle{cas-model2-names}

\bibliography{cas-refs}

@article{ingster2014signal,
  title={Signal detection for inverse problems in a multidimensional framework},
  author={Ingster, Yu and Laurent, B{\'e}atrice and Marteau, Cl{\'e}ment},
  journal={Mathematical Methods of Statistics},
  volume={23},
  number={4},
  pages={279--305},
  year={2014},
  publisher={Springer}
}

@article{marteau2015unified,
  title={A unified treatment for non-asymptotic and asymptotic approaches to minimax signal detection},
  author={Marteau, C and Sapatinas, T},
  journal={Statistics Surveys},
  volume={9},
  pages={253--297},
  year={2015}
}

@article{kroll2019rate,
  title={Rate optimal estimation of quadratic functionals in inverse problems with partially unknown operator and application to testing problems},
  author={Kroll, Martin},
  journal={ESAIM: Probability and Statistics},
  volume={23},
  pages={524--551},
  year={2019},
  publisher={EDP Sciences}
}

@article{kennedy2001bayesian,
  title={Bayesian calibration of computer models},
  author={Kennedy, Marc C and O'Hagan, Anthony},
  journal={Journal of the Royal Statistical Society: Series B (Statistical Methodology)},
  volume={63},
  number={3},
  pages={425--464},
  year={2001},
  publisher={Wiley Online Library}
}

@article{judd2004indistinguishable,
  title={Indistinguishable states II: The imperfect model scenario},
  author={Judd, Kevin and Smith, Leonard A},
  journal={Physica D: nonlinear phenomena},
  volume={196},
  number={3-4},
  pages={224--242},
  year={2004},
  publisher={Elsevier}
}

@article{berger2019statistical,
  title={On the statistical formalism of uncertainty quantification},
  author={Berger, James O and Smith, Leonard A},
  journal={Annual review of statistics and its application},
  volume={6},
  number={1},
  pages={433--460},
  year={2019},
  publisher={Annual Reviews}
}

@article{boulakia2013stability,
  title={Stability estimates for a Robin coefficient in the two-dimensional Stokes system},
  author={Boulakia, M and Egloffe, AC and Grandmont, C},
  journal={Mathematical Control and Related Fields},
  volume={3},
  number={1},
  pages={21--49},
  year={2013}
}

@article{arthern2010initialization,
  title={Initialization of ice-sheet forecasts viewed as an inverse Robin problem},
  author={Arthern, Robert J and Gudmundsson, G Hilmar},
  journal={Journal of Glaciology},
  volume={56},
  number={197},
  pages={527--533},
  year={2010},
  publisher={Cambridge University Press}
}

@article{martin2014adjoint,
  title={Adjoint accuracy for the full Stokes ice flow model: limits to the transmission of basal friction variability to the surface},
  author={Martin, Nathan and Monnier, J{\'e}r{\^o}me},
  journal={The Cryosphere},
  volume={8},
  number={2},
  pages={721--741},
  year={2014},
  publisher={Copernicus Publications G{\"o}ttingen, Germany}
}

@Article{wolovick2023regularization,
AUTHOR = {Wolovick, M. and Humbert, A. and Kleiner, T. and R\"uckamp, M.},
TITLE = {Regularization and L-curves in ice sheet inverse models: a case study in the Filchner--Ronne catchment},
JOURNAL = {The Cryosphere},
VOLUME = {17},
YEAR = {2023},
NUMBER = {12},
PAGES = {5027--5060},
URL = {https://tc.copernicus.org/articles/17/5027/2023/},
DOI = {10.5194/tc-17-5027-2023}
}

@book{morozov2012methods,
  title={Methods for solving incorrectly posed problems},
  author={Morozov, Vladimir Alekseevich},
  year={2012},
  publisher={Springer Science \& Business Media}
}

@article{ramdas2025hypothesis,
  title={Hypothesis testing with e-values},
  author={Ramdas, Aaditya and Wang, Ruodu},
  journal={Foundations and Trends in Statistics},
  volume={1},
  number={1-2},
  pages={1--390},
  year={2025},
  publisher={Emerald Publishing Limited}
}

@article{grunwald2024safe,
    author = {Grünwald, Peter and de Heide, Rianne and Koolen, Wouter},
    title = {Safe testing},
    journal = {Journal of the Royal Statistical Society Series B: Statistical Methodology},
    volume = {86},
    number = {5},
    pages = {1091-1128},
    year = {2024},
    month = {11},
    doi = {10.1093/jrsssb/qkae011},
}

@article{obrien1979multiple,
  title={A multiple testing procedure for clinical trials},
  author={O'Brien, Peter C and Fleming, Thomas R},
  journal={Biometrics},
  pages={549--556},
  year={1979},
  publisher={JSTOR}
}

@article{pocock1977group,
  title={Group sequential methods in the design and analysis of clinical trials},
  author={Pocock, Stuart J},
  journal={Biometrika},
  volume={64},
  number={2},
  pages={191--199},
  year={1977},
  publisher={Oxford University Press}
}

@book{jennison1999group,
  title={Group sequential methods with applications to clinical trials},
  author={Jennison, Christopher and Turnbull, Bruce W},
  year={1999},
  publisher={CRC press}
}

@article{gelman1996posterior,
  title={Posterior predictive assessment of model fitness via realized discrepancies},
  author={Gelman, Andrew and Meng, Xiao-Li and Stern, Hal},
  journal={Statistica sinica},
  pages={733--760},
  year={1996},
  publisher={JSTOR}
}

@article{vovk2021values,
  title={E-values: Calibration, combination and applications},
  author={Vovk, Vladimir and Wang, Ruodu},
  journal={The Annals of Statistics},
  volume={49},
  number={3},
  pages={1736--1754},
  year={2021},
  publisher={Institute of Mathematical Statistics}
}

@book{ville1939etude,
  title={Etude critique de la notion de collectif},
  author={Ville, Jean},
  volume={3},
  year={1939},
  publisher={Gauthier-Villars Paris}
}

@article{ramdas2023game,
  title={Game-theoretic statistics and safe anytime-valid inference},
  author={Ramdas, Aaditya and Gr{\"u}nwald, Peter and Vovk, Vladimir and Shafer, Glenn},
  journal={Statistical Science},
  volume={38},
  number={4},
  pages={576--601},
  year={2023},
  publisher={Institute of Mathematical Statistics}
}

@article{grunwald2024beyond,
  title={Beyond Neyman--Pearson: E-values enable hypothesis testing with a data-driven alpha},
  author={Gr{\"u}nwald, Peter D},
  journal={Proceedings of the National Academy of Sciences},
  volume={121},
  number={39},
  pages={e2302098121},
  year={2024},
  publisher={National Academy of Sciences}
}

@article{howard2021time,
  title={Time-uniform, nonparametric, nonasymptotic confidence sequences},
  author={Howard, Steven R and Ramdas, Aaditya and McAuliffe, Jon and Sekhon, Jasjeet},
  journal={The Annals of Statistics},
  volume={49},
  number={2},
  pages={1055--1080},
  year={2021},
  publisher={JSTOR}
}

@article{shapero2021icepack,
  title={icepack: A new glacier flow modeling package in Python, version 1.0},
  author={Shapero, Daniel R and Badgeley, Jessica A and Hoffman, Andrew O and Joughin, Ian R},
  journal={Geoscientific Model Development},
  volume={14},
  number={7},
  pages={4593--4616},
  year={2021},
  publisher={Copernicus Publications G{\"o}ttingen, Germany}
}

@article{recinos2025mapping,
  title={Mapping ice stream sensitivity in the Amundsen Sector to uncertainty in ice velocity observations},
  author={Recinos, Beatriz and Goldberg, Daniel and Gourmelen, Noel and Maddison, James R},
  journal={Geophysical Research Letters},
  volume={52},
  number={17},
  pages={e2025GL117666},
  year={2025},
  publisher={Wiley Online Library}
}

@article{johari2022always,
  title={Always valid inference: Continuous monitoring of a/b tests},
  author={Johari, Ramesh and Koomen, Pete and Pekelis, Leonid and Walsh, David},
  journal={Operations Research},
  volume={70},
  number={3},
  pages={1806--1821},
  year={2022},
  publisher={INFORMS}
}

@article{ter2020anytime,
  title={The anytime-valid logrank test: Error control under continuous monitoring with unlimited horizon},
  author={Ter Schure, Judith and P{\'e}rez-Ortiz, Muriel F and Ly, Alexander and Grunwald, P},
  journal={arXiv preprint arXiv:2011.06931},
  year={2020}
}

@article{henzi2022valid,
  title={Valid sequential inference on probability forecast performance},
  author={Henzi, Alexander and Ziegel, Johanna F},
  journal={Biometrika},
  volume={109},
  number={3},
  pages={647--663},
  year={2022},
  publisher={Oxford University Press}
}

@article{stuart2010inverse,
  title={Inverse problems: a Bayesian perspective},
  author={Stuart, Andrew M},
  journal={Acta numerica},
  volume={19},
  pages={451--559},
  year={2010},
  publisher={Cambridge University Press}
}

@book{engl1996regularization,
  title={Regularization of inverse problems},
  author={Engl, Heinz Werner and Hanke, Martin and Neubauer, Andreas},
  volume={375},
  year={1996},
  publisher={Springer Science \& Business Media}
}

@book{alifanov2012inverse,
  title={Inverse heat transfer problems},
  author={Alifanov, Oleg M},
  year={2012},
  publisher={Springer Science \& Business Media}
}

@article{borcea2002electrical,
  title={Electrical impedance tomography},
  author={Borcea, Liliana},
  journal={Inverse problems},
  volume={18},
  number={6},
  pages={R99--R136},
  year={2002}
}

@book{vogel2002computational,
  title={Computational methods for inverse problems},
  author={Vogel, Curtis R},
  year={2002},
  publisher={SIAM}
}

@inproceedings{podkopaev2022tracking,
  title={Tracking the risk of a deployed model and detecting harmful distribution shifts},
  author={Podkopaev, Aleksandr and Ramdas, Aaditya},
  booktitle={Tenth International Conference on Learning Representations},
  year={2022},
}

@article{cover1991universal,
  title={Universal portfolios},
  author={Cover, Thomas M},
  journal={Mathematical finance},
  volume={1},
  number={1},
  pages={1--29},
  year={1991},
  publisher={Wiley Online Library}
}

@article{alnaes2015fenics,
  title={The FEniCS project version 1.5},
  author={Aln{\ae}s, Martin and Blechta, Jan and Hake, Johan and Johansson, August and Kehlet, Benjamin and Logg, Anders and Richardson, Chris and Ring, Johannes and Rognes, Marie E and Wells, Garth N},
  journal={Archive of numerical software},
  volume={3},
  number={100},
  year={2015}
}

@article{duffin2021statistical,
  title={Statistical finite elements for misspecified models},
  author={Duffin, Connor and Cripps, Edward and Stemler, Thomas and Girolami, Mark},
  journal={Proceedings of the National Academy of Sciences},
  volume={118},
  number={2},
  pages={e2015006118},
  year={2021},
  publisher={National Academy of Sciences}
}

@article{wasserman2020universal,
  title={Universal inference},
  author={Wasserman, Larry and Ramdas, Aaditya and Balakrishnan, Sivaraman},
  journal={Proceedings of the National Academy of Sciences},
  volume={117},
  number={29},
  pages={16880--16890},
  year={2020},
  publisher={National Academy of Sciences}
}

@article{rahimi2007random,
  title={Random features for large-scale kernel machines},
  author={Rahimi, Ali and Recht, Benjamin},
  journal={Advances in neural information processing systems},
  volume={20},
  year={2007}
}

@article{jin2010numerical,
  title={Numerical estimation of the Robin coefficient in a stationary diffusion equation},
  author={Jin, Bangti and Zou, Jun},
  journal={IMA Journal of Numerical Analysis},
  volume={30},
  number={3},
  pages={677--701},
  year={2010},
  publisher={Oxford University Press}
}

@article{petra2012inexact,
  title={An inexact Gauss-Newton method for inversion of basal sliding and rheology parameters in a nonlinear Stokes ice sheet model},
  author={Petra, Noemi and Zhu, Hongyu and Stadler, Georg and Hughes, Thomas JR and Ghattas, Omar},
  journal={Journal of Glaciology},
  volume={58},
  number={211},
  pages={889--903},
  year={2012},
  publisher={Cambridge University Press}
}

@article{waudby2025universal,
  title={Universal log-optimality for general classes of e-processes and sequential hypothesis tests},
  author={Waudby-Smith, Ian and Sandoval, Ricardo and Jordan, Michael I},
  journal={arXiv preprint arXiv:2504.02818},
  year={2025}
}

@article{rasmussen2024bayesian,
  title={The Bayesian approach to inverse Robin problems},
  author={Rasmussen, Aksel K and Seizilles, Fanny and Girolami, Mark and Kazlauskaite, Ieva},
  journal={SIAM/ASA Journal on Uncertainty Quantification},
  volume={12},
  number={3},
  pages={1050--1084},
  year={2024},
  publisher={SIAM}
}

@article{brynjarsdottir2014learning,
  title={Learning about physical parameters: The importance of model discrepancy},
  author={Brynjarsd{\'o}ttir, Jenn{\`y} and O'Hagan, Anthony},
  journal={Inverse problems},
  volume={30},
  number={11},
  pages={114007},
  year={2014},
  publisher={IOP Publishing}
}

@article{rathgeber2016firedrake,
  title={Firedrake: automating the finite element method by composing abstractions},
  author={Rathgeber, Florian and Ham, David A and Mitchell, Lawrence and Lange, Michael and Luporini, Fabio and McRae, Andrew TT and Bercea, Gheorghe-Teodor and Markall, Graham R and Kelly, Paul HJ},
  journal={ACM Transactions on Mathematical Software (TOMS)},
  volume={43},
  number={3},
  pages={1--27},
  year={2016},
  publisher={ACM New York, NY, USA}
}

@article{shafer2021testing,
    author = {Shafer, Glenn},
    title = {Testing by Betting: A Strategy for Statistical and Scientific Communication},
    journal = {Journal of the Royal Statistical Society Series A: Statistics in Society},
    volume = {184},
    number = {2},
    pages = {407-431},
    year = {2021},
    month = {04}
}

\newpage
\appendix

\section{Misspecification geometry}\label{sec:si-geometry}

\begin{figure}[htbp]
  \centering
  \includegraphics[width=0.8\textwidth]{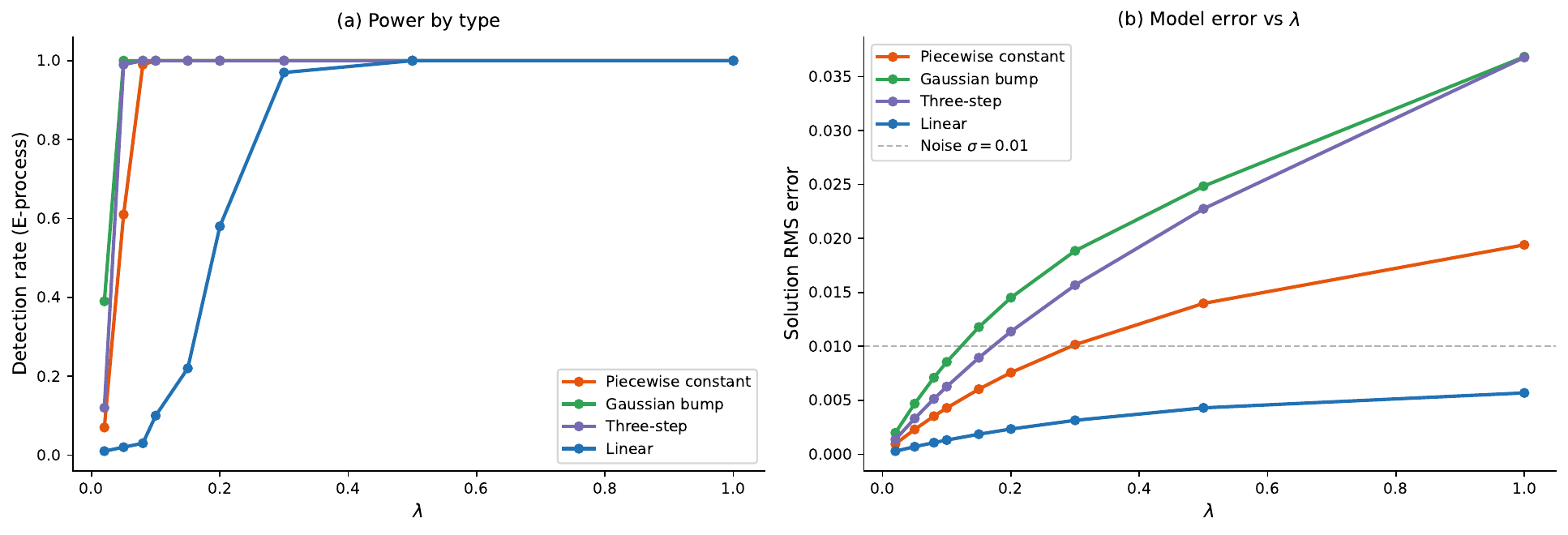}
  \caption{Detection power and solution-space model error across the four misspecification types. Left: e-process detection rate versus $\lambda$. Right: RMS of the noise-free solution error versus $\lambda$, with the observation-noise level $\sigma=0.01$ shown dashed. Differences in power at comparable RMS reflect how residual structure projects onto the expert bank.}
  \label{fig:overlay}
\end{figure}

Detection difficulty depends not only on model-error magnitude but also on how the error projects onto the expert bank. Fig.~\ref{fig:overlay} compares the e-process detection rate and the solution-space RMS error as $\lambda$ varies across all four misspecification types. Detection generally becomes easier as the RMS error grows, but the power curves differ across types even at comparable RMS levels, showing that magnitude alone does not determine detectability. This is consistent with their spatial structure (Fig.~\ref{fig:diffusivity}): the bump is localised, the linear alternative is diffuse, and the piecewise and three-step alternatives distribute energy across several harmonics. How these residual patterns project onto the expert bank therefore helps explain the differences in detection power beyond their RMS magnitude.

The four alternative diffusivity functions $\theta_{\mathrm{alt}}(x)$ used in the Poisson experiments are:
\begin{itemize}
\setlength{\itemsep}{1pt}
  \item Piecewise constant: $2 \cdot \mathbf{1}_{x<0.5} + 8 \cdot \mathbf{1}_{x \geq 0.5}$
  \item Gaussian bump: $3 + 5\exp(-(x-0.7)^2/0.02)$
  \item Three-step: $2 \cdot \mathbf{1}_{x<1/3} + 6 \cdot \mathbf{1}_{1/3 \leq x < 2/3} + 3 \cdot \mathbf{1}_{x \geq 2/3}$
  \item Linear: $1 + 8x$
\end{itemize}
The interpolated diffusivity is $\theta_\lambda(x) = (1-\lambda)\theta_{\mathrm{null}}(x) + \lambda\,\theta_{\mathrm{alt}}(x)$, where $\theta_{\mathrm{null}}(x) = \exp(1.1x + 1.5x^3)$.

\section{Stopping time comparison}\label{sec:si-stop-comparison}

\begin{figure}[htbp]
  \centering
  \includegraphics[width=0.8\textwidth]{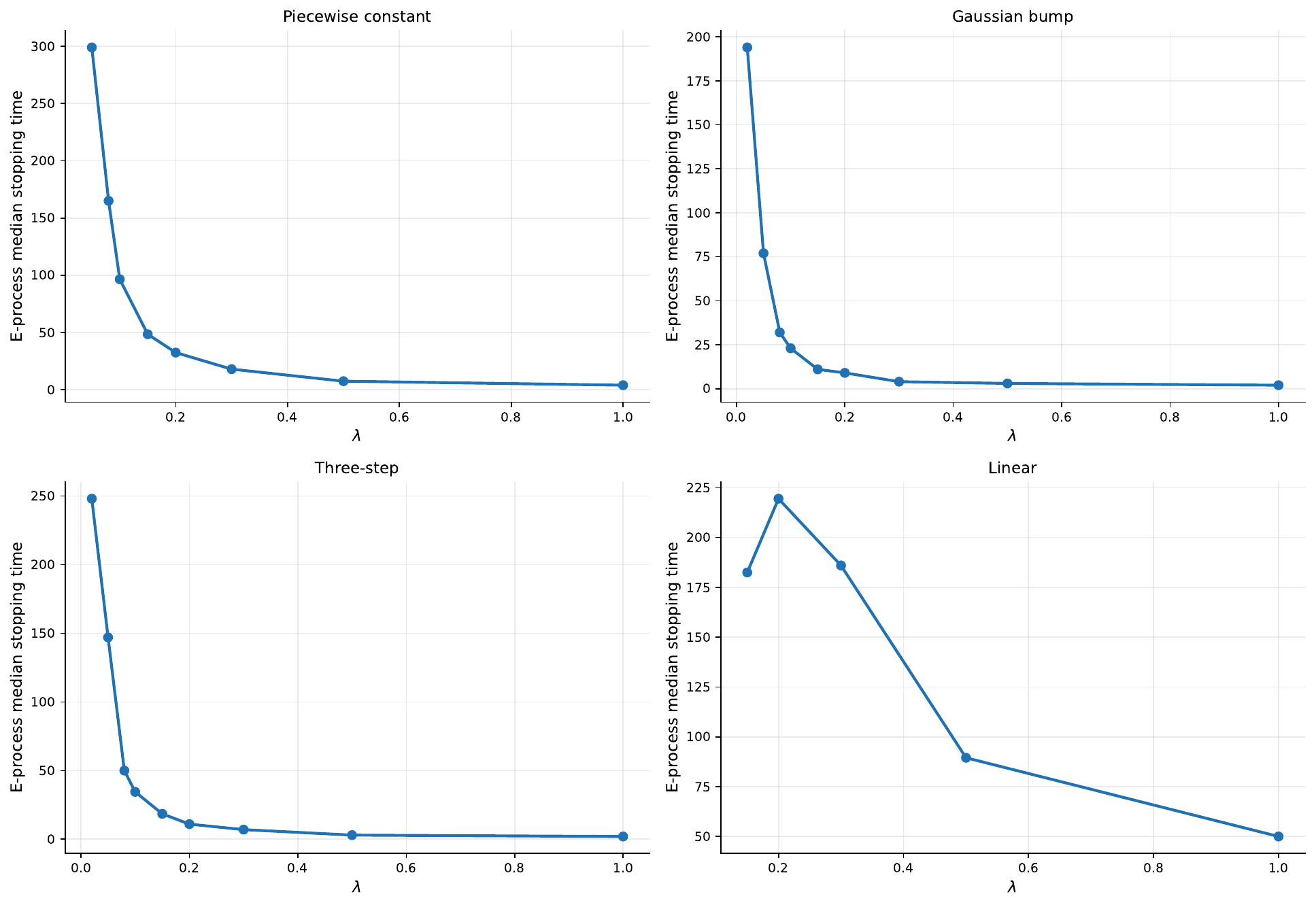}
  \caption{E-process median stopping time vs.\ $\lambda$ for the four misspecification types, where detection rate $> 10\%$. Stopping time falls monotonically with $\lambda$ for the bump, piecewise, and three-step cases; the subnoise linear case is non-monotonic and slower. The full comparison with the magnitude-based tests is in Table~\ref{tab:si-full-power}.}
  \label{fig:stopping}
\end{figure}

Fig.~\ref{fig:stopping} shows the e-process median stopping time as a
function of misspecification strength $\lambda$. Among the displayed
cells, which are restricted to detection rates above $10\%$, stopping
time decreases monotonically with $\lambda$ for the bump, piecewise,
and three-step alternatives, reaching single-digit observation counts
at $\lambda=1$. The linear case is non-monotonic and slower: at
$\lambda=1.0$ the e-process detects in $\PoisLinearLamOneEMed$
observations. The full E/Bonferroni/Pocock/OBF comparison appears in Table~\ref{tab:si-full-power}.

The key advantage of the e-process over Bonferroni is not speed but \emph{detection rate}: the e-process's terminal evidence $E_T$ is order-invariant, so its detection rate is approximately ordering-insensitive (exactly so for rejection by $E_T \geq 1/\alpha$; approximately so for the sequential crossing rule we use, where the order affects whether the running maximum reaches the threshold; Appendix~\ref{sec:si-ordering} shows this effect is empirically small in our experiments). The Bonferroni sequential test, by contrast, is highly sensitive to the order in which observations arrive. With randomly ordered observations, the Bonferroni test loses substantial power in the transition region (\emph{e.g.}, piecewise $\lambda = 0.10$: $\PoisPieceLamPoneBonfDet\%$ vs.\ $\PoisPieceLamPoneEDet\%$ for the e-process).

\section{QoI error scan}\label{sec:si-qoi-scan}

Table~\ref{tab:qoi} reports the deterministic QoI error and Morozov status across all misspecification types and strengths. The dangerous zone (where Morozov accepts but QoI error exceeds 5\%) covers large portions of the parameter space, especially for the linear and three-step cases.

\begin{table}[htbp]
\centering
\caption{QoI errors across misspecification types and strengths. ``ok'' means Morozov accepts the model; ``FLAG'' means Morozov rejects (decision rule with $\tau = 1.5$).}
\label{tab:qoi}
\small
\begin{tabular}{@{}llcccrr@{}}
\toprule
Type & $\lambda$ & RMS & Flux & Avg $u$ & $u(0.95)$ & Mrzv \\
\midrule
Linear & 0.10 & .001 & 0.4\% & 2.4\% & 4.7\% & ok \\
 & 0.20 & .002 & 0.8\% & 4.2\% & 8.3\% & ok \\
 & 0.50 & .004 & 2.0\% & 7.0\% & 14.3\% & ok \\
 & 1.00 & .005 & 3.9\% & 7.8\% & 15.3\% & ok \\
\addlinespace
Gaussian bump & 0.05 & .004 & 1.3\% & 2.9\% & 6.3\% & ok \\
 & 0.10 & .008 & 2.6\% & 5.4\% & 12.0\% & ok \\
 & 0.20 & .014 & 4.9\% & 9.1\% & 21.5\% & FLAG \\
\addlinespace
Piecewise constant & 0.05 & .002 & 0.4\% & 1.4\% & 3.2\% & ok \\
 & 0.10 & .004 & 0.8\% & 2.5\% & 5.9\% & ok \\
 & 0.20 & .007 & 1.8\% & 4.1\% & 9.9\% & ok \\
 & 0.50 & .013 & 4.5\% & 6.5\% & 15.9\% & FLAG \\
\addlinespace
Three-step & 0.05 & .003 & 0.3\% & 4.2\% & 7.6\% & ok \\
 & 0.10 & .006 & 0.7\% & 7.7\% & 14.3\% & ok \\
 & 0.15 & .008 & 1.0\% & 10.7\% & 20.1\% & ok \\
 & 0.20 & .011 & 1.5\% & 13.1\% & 25.2\% & FLAG \\
\addlinespace
\bottomrule
\end{tabular}
\end{table}

\section{Stokes inverse problem: additional details}\label{sec:si-stokes}

The Stokes flow problem is solved on $[0,1] \times [0, 0.2]$ with Taylor--Hood (P2/P1) finite elements on a $150 \times 30$ triangle mesh (approximately 40,000 degrees of freedom). The physical parameters are: density $\rho = 1$, gravity components $g_x = g_y = 5$, and surface stress $\tau(x) = 10(\sin(12\pi x) + 1)$.

The basal drag is parameterised as
$\beta(x)=\exp(\theta_0+\sum_{k=1}^{K-1}\theta_k\phi_k(x))$,
where $\phi_k(x)=\cos(2\pi jx)$ for odd $k$ and $\phi_k(x)=\sin(2\pi jx)$ for even $k$, with $j=1+\lfloor(k-1)/2\rfloor$. The exponentiation ensures positivity. The true parameters are \stokesThetaTruthFormatted\ with $K = 4$ coefficients.

The misspecified model uses $K=3$ coefficients. The oracle fit is obtained by Nelder--Mead minimisation of
\[
\sum_{i=1}^{N}
\left[
  \bigl(u_x^{\mathrm{truth}}(x_i)-u_{x,\theta}(x_i)\bigr)^2
  +
  \bigl(u_y^{\mathrm{truth}}(x_i)-u_{y,\theta}(x_i)\bigr)^2
\right],
\]
using both surface-velocity components. The noisy-fit validation uses the analogous objective with independent Gaussian noise of standard deviation $\sigma$ added to each component. All diagnostics reported in the paper are subsequently applied only to the $u_x$ residual channel. The fitted parameters \stokesThetaOracleKThreeFormatted\ absorb part of the omitted $\phi_3(x)=\cos(4\pi x)$ mode into the first three coefficients, producing a surface velocity field that is visually nearly indistinguishable from truth while leaving spatially structured $u_x$ residuals at $\mathrm{RMS}/\sigma\approx\stokesRmsObsOverSigmaMean$.

The mechanism behind this diagnostic failure is the forward operator's smoothing: the Stokes equations map basal drag to surface velocity through an elliptic operator that attenuates high-frequency perturbations. Omission of the second cosine harmonic, $\cos(4\pi x)$, from the parameterisation of $\log\beta(x)$ produces a surface-velocity error that is substantially smaller in magnitude, pushing the residual RMS below Morozov's threshold while preserving the spatial structure that the e-process detects.

The practical consequence is a decoupling between observation-space fit quality and parameter-space accuracy. The $K = 3$ model reproduces surface velocities to \stokesUxRelErrInteriorKThree, but the pointwise basal drag $\beta(x)$ has up to \stokesBetaPointwiseMaxKThree\ error (mean \stokesBetaPointwiseMeanKThree). Regional averages in the central domain ($x \in [\stokesRegionalIntervalLo, \stokesRegionalIntervalHi]$) are wrong by \stokesBetaRegionalMeanKThree\ on average. The location of minimum drag, the point most vulnerable to grounding line retreat in glaciology, shifts from $x = \stokesBetaArgminTruth$ to $x = \stokesBetaArgminKThree$. In an ice-sheet projection, this would mean predicting destabilisation at the wrong location.

\section{\textit{icepack} ice-stream inverse problem: additional details}\label{sec:si-icepack}

\subsection{Forward model and geometry}

The third example uses \textit{icepack}~\citep{shapero2021icepack}, an open-source finite-element ice-flow model built on Firedrake. We adapt the package's standard synthetic ice-stream tutorial by modifying the bed and surface geometry so the entire domain is grounded, eliminating the floating shelf and grounding-line dynamics that are not the subject of this experiment. All modifications are confined to the input data: the forward operator (the shallow shelf approximation with Weertman friction), the solver, and the function-space construction are used as supplied by \textit{icepack}.

The domain is rectangular, $[0, L_x] \times [0, L_y]$ with $L_x = 50$~km and $L_y = 12$~km, discretised on a $48 \times 32$ rectangular mesh. The bed elevation drops linearly from $b_{\text{in}} = 400$~m to $b_{\text{out}} = 100$~m along the flow direction, and the surface from $s_{\text{in}} = 850$~m to $s_{\text{out}} = 200$~m, giving ice thicknesses of $450$~m at the inflow and $100$~m at the terminus. With both bed and surface above sea level, the effective-pressure factor in the friction law equals unity throughout, and there is no flotation. Boundary conditions follow the \textit{icepack} tutorial: Dirichlet inflow at $x = 0$, lateral free-slip at $y = 0$ and $y = L_y$, and natural (stress-free) terminus at $x = L_x$. Ice rheology uses Glen's flow law with $n = 3$ and a fluidity computed at $-13^\circ$C.

The basal friction parameterisation is a related truncated-Fourier construction used in the Stokes case (Appendix~\ref{sec:si-stokes}):
\[
C(x) = C_0 \exp\!\left(\theta_0 + \sum_{k=1}^{K-1}\theta_k \phi_k(x/L_x)\right),
\]
with $\phi_k = \cos(k\pi s)$ for odd $k$ and $\phi_k = \sin(k\pi s)$ for even $k$, where $s = x/L_x$. The baseline $C_0 = 0.40 \cdot \tau_D / u_{\text{in}}^{1/m}$ where $\tau_D = \rho_I g h_{\text{in}}|\partial s/\partial x|$ is the driving stress at the inflow, $u_{\text{in}} = 20$~m/yr is a reference inflow speed, and $m = 3$ is the Weertman exponent. The factor $0.40$ (rather than the \textit{icepack} tutorial's $0.95$) places the ice in a sliding-dominated regime where membrane stresses carry a substantial fraction of the driving stress, ensuring that variations in $C(x)$ produce visible velocity variations in the surface field.

True parameters: $\theta_{\text{truth}} = (-0.05, 0.15, 0.10, 0.40)$ with $K = 4$. The amplitudes are smaller than the Stokes case ($(-0.6, 1.7, 0.3, 0.5)$) because $C_0$ enters dimensionally and the \textit{icepack} geometry has different scales; we tuned $\theta_{\text{truth}}$ so that the resulting velocity field is in the realistic ice-stream range ($20$--$144$~m/yr, with mean $73$~m/yr in the observation region) and shows visible structure from the friction parameterisation. The misspecified model uses $K = 3$ coefficients.

We do \emph{not} run a prognostic spin-up. The inverse problem is defined by the diagnostic forward operator $C \mapsto u$ on a fixed geometry $(h_0, s_0)$, mirroring the Stokes example. A spin-up would couple thickness evolution to the friction field and confound the inverse-problem identification.

\subsection{Observations, fit, and diagnostic comparison}

Observations are the $x$-component of surface velocity sampled at $N = 200$ random $(x, y)$ locations in the well-resolved interior, $x \in [5, 45]$~km. The interval excludes the inflow boundary layer and the terminus stress ramp, where the natural BCs introduce edge effects unrelated to the friction parameterisation. Velocities at observation points span $39$--$106$~m/yr with mean $73$~m/yr. The same sampled observation locations are used across Monte Carlo realisations.

For orientation, we first compute a reference misspecified $K = 3$ fit by Nelder--Mead minimisation of the sum of squared residuals against the noise-free truth observations, with up to 200 iterations and convergence tolerances $\Delta x \leq 10^{-3}$, $\Delta f \leq 10^{-1}$. Each function evaluation requires one \textit{icepack} diagnostic solve (a few seconds on a single core); convergence is reached in 98 iterations and 177 evaluations. The fitted parameters are $\hat\theta = (-0.044, 0.96, -0.80)$, with the lower harmonics shifted by factors of 6--8 relative to the truth's first three coefficients to absorb part of the missing harmonic. This reference fit is used to describe the deterministic model-error structure: $u_{\text{truth}} - u_{\text{fit}}$ has $\mathrm{RMS} = 1.19$~m/yr at observation points, mean error zero, and maximum pointwise error $4.5$~m/yr.

The reported diagnostic rates, however, use noisy refits rather than this single noise-free reference fit. In each of 100 Monte Carlo realisations, independent $N(0,\sigma^2)$ noise is added to the truth observations, the misspecified $K = 3$ model is refit by Nelder--Mead, and diagnostics are applied to the resulting residuals. The $K = 4$ null calibration is generated analogously from 100 noisy refits of the correctly specified model. These \textit{icepack} results therefore evaluate the empirical full fit-and-test pipeline; the formal e-process martingale guarantee applies conditionally once the fitted model is fixed.

We choose the observation noise $\sigma = 1.3$~m/yr so that the
residual $\mathrm{RMS}/\sigma$ matches the Stokes example: across the
100 noisy $K=3$ refits, $\mathrm{RMS}/\sigma$ has mean $\IceERMS$ and
range $[1.19,1.52]$. This is comparable to satellite ice-velocity precision in slow-flow regions and to GPS-based in-situ measurements~\citep{recinos2025mapping}.

The expert bank uses the same construction as the other examples: $K = 156$ experts from $5$ Fourier frequencies (sin and cos), $3$ polynomial degrees, each at $12$ signed amplitudes spanning $0.5\sigma$ to $8\sigma$ in magnitude. The amplitude magnitudes scale with $\sigma$: $\pm\{0.65, 1.3, 2.6, 3.9, 6.5, 10.4\}$ for $\sigma = 1.3$. Other configuration is identical to the Poisson and Stokes cases.

Over 100 independent noisy $K = 3$ refits: Morozov accepts in $\IceMrzvAcceptPct\%$; the fixed-sample $\chi^2$ test rejects in $\IceFCDet\%$ (full data); the batch Fourier test rejects in $\IceBFDet\%$ (full data); the e-process rejects in $\IceEDet\%$ with median stopping time $\IceEMed$ observations (mean $\IceEMean$). Pocock $\alpha$-spending rejects $\IcePocockDet\%$ at median $\IcePocockMed$ (mean $\IcePocockMean$); OBF at median $\IceOBFMed$ (mean $\IceOBFMean$). Null rejection rates from 100 independent noisy $K = 4$ refits are: Morozov $\IceMrzvNullRate$, batch Fourier $\IceBFNullRate$, e-process $\IceENullRate$, Pocock $\IcePocockNullRate$, OBF $\IceOBFNullRate$, all at or below the nominal $\alpha = 0.05$. The Bonferroni sequential $\chi^2$ test rejects $\IceBonfDet\%$ with median stopping time $\IceBonfMed$ (mean $\IceBonfMean$) but has a null rejection rate of $\IceBonfNullRate$, confirming the anti-conservatism we report for this method elsewhere (Appendix~\ref{sec:si-calibration}); we omit it from Table~\ref{tab:icepack}.

\subsection{Diagnosis: residual structure}

The portfolio wealth distribution concentrates sharply: across 100 MC runs, the mean wealth on the $\cos(5\pi x/L_x)$ expert at amplitude $1.3 = 1.0\sigma$ is $0.91$, with a further $0.09$ on the same expert at amplitude $0.65 = 0.5\sigma$. All other experts receive negligible weight. During refitting, the retained coefficients $\theta_1$ and $\theta_2$ shift substantially to absorb part of the omitted mode's effect in observation space; after propagation through the nonlinear friction law and forward operator, the remaining velocity residual is empirically dominated by $\cos(5\pi x/L_x)$. This is consistent with the residual scatter (Fig.~\ref{fig:icepack}b), which shows two and a half visible oscillations of the noise-free model error across the interior.

\subsection{Decision-relevant consequence}

The accepted $K = 3$ model produces inferred basal friction $\hat C(x)$ that deviates from truth by up to $\IceBasalFrictionErrMaxPct\%$ pointwise, with the largest error at the inflow boundary ($x = 0$, $\hat C/C_{\text{truth}} \approx 1.5$). Interior pointwise error is roughly $10\%$. We compute two quantities of interest:

\textit{Interior surface velocity}: at $x = 25$~km, $u_{\text{truth}} = 82.3$~m/yr and $u_{\text{fit}} = 81.9$~m/yr, a relative error of $0.5\%$. At $x = 42$~km, $u_{\text{truth}} = 41.1$~m/yr and $u_{\text{fit}} = 40.5$~m/yr, $1.5\%$. The fit reproduces interior velocities to within typical observation noise.

\textit{Terminus ice flux}: at $x = L_x$, $u_{\text{truth}} = 144$~m/yr and $u_{\text{fit}} = 196$~m/yr, a relative error of $36.2\%$ uniform across the terminus cross-section (the velocity field is nearly $y$-independent because the friction parameterisation is a function of $x$ only). The integrated flux through the terminus is correspondingly about \(36\%\) higher under the accepted model. In an operational projection workflow, such a flux bias could materially bias the inferred catchment mass-loss contribution, although translating it into sea-level contribution would require the full geometry, density conversion, and transient mass-balance model.

The mechanism is the same as the Poisson example's near-boundary failure: the K=3 fit's parameter error is concentrated at the boundary, and the velocity field carries this error through the elliptic operator to where it concentrates again at the opposite boundary, where ice flux is computed. Surface observations in the interior give no warning of this: interior velocities are reproduced almost exactly.

\section{Expert design and the discrepancy-modelling bridge}\label{sec:si-expert-design}

The Fourier–polynomial expert set works well in our experiments for two reasons. First, the growth-rate theory of e-processes (\S\ref{sec:method-portfolio}) characterises which experts profit under the alternative; this is not specific to a basis but determines what makes a basis function effective. Second, for a squared-exponential GP discrepancy on a regular interval, the leading Karhunen–Loève eigenfunctions are approximately Fourier modes, motivating the choice of basis. The match depends on the domain, boundary conditions, kernel, and measure but it provides a useful interpretive bridge between the expert design and the Bayesian framework of~\citet{kennedy2001bayesian}.

\paragraph{KL connection.} In the~\citet{kennedy2001bayesian} framework, $\delta(x) \sim \mathrm{GP}(0, k_\ell)$ with, say, a squared-exponential kernel of length scale $\ell$. For a periodic squared-exponential kernel on the torus, the Fourier basis diagonalises the covariance operator and the eigenvalues decay like the spectral density, approximately $\exp(-c\,j^2\ell^2)$ for a constant $c$ fixed by the frequency convention. For the same kernel restricted to $[0,1]$ with non-periodic boundary behaviour, the Karhunen--Lo\`eve eigenfunctions are not exactly sine and cosine, though low-frequency Fourier modes remain a useful approximation in the interior. The Fourier bank is therefore best read as a smooth low-frequency dictionary motivated by the spectral decay of common GP kernels, rather than as the exact KL basis. The amplitude grid plays the role of the prior on the discrepancy scale $\tau$: the eigenvalue $\lambda_j$ sets the prior scale of the $j$-th coefficient, $\lambda_j^{1/2}\tau$. This is distinct from the testing-optimal amplitude: for a \emph{realised} discrepancy $\mu^*$, the expert amplitude that maximises expected log-likelihood growth along mode $\phi_j$ is its projection coefficient $a_j^* = \langle\mu^*, \phi_j\rangle_\nu / \langle\phi_j, \phi_j\rangle_\nu$ (\S\ref{sec:method-portfolio}), which depends on the actual discrepancy rather than on the prior.

In the statFEM framework of~\citet{duffin2021statistical}, the stochastic forcing $\xi_\theta$ plays an analogous role: its covariance kernel determines which spatial patterns the discrepancy can produce, and the KL eigenfunctions of that kernel would be the natural experts. For non-stationary or non-smooth kernels (e.g.\ Matérn $1/2$), the natural expert basis differs from Fourier; the effectiveness of Fourier--polynomial experts in our examples reflects the smoothness of the discrepancies we encounter, not a property of the framework.

\paragraph{Connection to neural discrepancy models.} The same connection extends to neural discrepancy models in a limited sense: a shallow network with sinusoidal activations contains the Fourier experts as special fixed-frequency units, and random Fourier features~\citep{rahimi2007random} approximate the feature representation of stationary Gaussian process kernels. The e-process portfolio weights can therefore be interpreted as a cheap frequency-selection diagnostic: they suggest which spatial scales carry discrepancy signal before a richer correction model is trained.

\section{Effect of observation ordering}\label{sec:si-ordering}

The Poisson and Stokes experiments in the main text randomly permute the
residuals independently within each Monte Carlo run. The
\textit{icepack} experiment instead uses one fixed cached observation
order. Both are pre-specified or independently randomised orderings;
neither is selected retrospectively after inspecting the residuals.

The e-process's type-I error guarantee holds for any pre-specified ordering, and more generally for any predictable ordering, where ``predictable'' means the choice of which observation to process at step $t$ is fixed in advance or determined by past residuals only, without looking at future ones. The guarantee does \emph{not} extend to retrospective orderings designed after observing all residuals to maximise the trajectory supremum. The final evidence $E_T = \sum_k \pi_k \exp(\sum_{t=1}^T \ell_{t,k})$ after all $T$ observations is order-invariant, since each cumulative log-likelihood ratio $L_{T,k}$ is a sum over all observations. A test that rejects when $E_T \geq 1/\alpha$ therefore has fully order-invariant detection rates. The sequential rule we actually use, rejecting when $E_t \geq 1/\alpha$ at any $t \leq T$, is \emph{not} order-invariant: the stopping time $\tau$, the supremum $\sup_{t \leq T} E_t$, and (for weak-signal trajectories) the binary rejection outcome all depend on the order in which residuals are presented. For fixed-fit experiments satisfying the residual null, the reported detection rate is an order-specific estimate of the anytime-valid detection probability. The \textit{icepack} ordering is predictable, but its same-data refitting qualification remains separate.

The Bonferroni sequential $\chi^2$ test does not share this property. It rejects at time $t$ if $p_t \leq \alpha/t$, where $p_t$ is computed from the cumulative statistic $S_t = \sum_{i=1}^t (r_i/\sigma)^2$ with $t$ degrees of freedom. Since the threshold $\alpha/t$ becomes stricter as $t$ grows, the test responds more readily to evidence that arrives early in the sequence. Under spatial ordering, certain misspecification types (e.g., Gaussian bump centred at $x = 0.7$) concentrate residual energy in a particular spatial region; if that region happens to be processed early, the Bonferroni test benefits. Under random ordering, this advantage is lost.

Table~\ref{tab:ordering} compares the two orderings on selected settings. Detection rates for the e-process are identical in these selected simulations. Bonferroni detection rates drop substantially under random ordering in the transition region, where the test's power is marginal. E-process stopping times remain order-dependent and move in a problem-specific direction: random ordering delays detection for the piecewise and bump alternatives, but accelerates it for the linear alternative. Bonferroni detection drops substantially for the piecewise and bump alternatives and slightly for the linear alternative.
\begin{table}[htbp]
\centering
\caption{Effect of observation ordering on detection rates and stopping times (100 Monte Carlo runs per case, default bank $K=156$, $\alpha=0.05$).}
\label{tab:ordering}
\small
\begin{tabular}{@{}llcccccc@{}}
\toprule
& & \multicolumn{3}{c}{Spatial order} & \multicolumn{3}{c}{Random order} \\
\cmidrule(lr){3-5}\cmidrule(lr){6-8}
Type & $\lambda$ & E det & B det & E stop & E det & B det & E stop \\
\midrule
Piecewise & 0.10 & 1.0 & 0.93 & 27 & 1.0 & 0.39 & 96.5 \\
Bump & 0.05 & 1.0 & 0.94 & 30 & 1.0 & 0.52 & 77 \\
Linear & 1.00 & 1.0 & 0.90 & 79 & 1.0 & 0.87 & 50 \\
\bottomrule
\end{tabular}
\end{table}

In applied settings where the analyst does not control the order of data acquisition, the order-invariance of $E_T$ means that the final-evidence-based detection rate reflects the intrinsic difficulty of the detection problem, not the accident of observation sequence.

\section{Sequential magnitude-based test calibration}\label{sec:si-calibration}
The sequential magnitude-based comparators in Tables~\ref{tab:reassurance} and~\ref{tab:icepack} adapt boundaries from classical group-sequential testing, whose standard derivations assume a Gaussian (Brownian information-time) joint law for the interim statistics. Our cumulative-$\chi^2$ statistics do not follow that law, and the naive $\alpha/t$ rule additionally overspends its budget ($\sum_t \alpha/t > \alpha$), so none of these rules retains exactly nominal type-I error. Table~\ref{tab:calibration} reports empirical false alarm rates from $200$ $H_0$ Monte Carlo runs at $T = 501$, the Poisson observation budget.

\begin{table}[htbp]
\centering
\caption{Empirical type-I rates of sequential and batch diagnostics at $T = 501$ ($n = 200$ $H_0$ Monte Carlo runs, $\alpha = 0.05$). The e-process is anytime-valid with exact control $\leq \alpha$ regardless of horizon; its empirical rate reflects the inequality's slack, not anti-conservatism.}
\label{tab:calibration}
\small
\begin{tabular}{@{}lc@{}}
\toprule
Test & Empirical type-I rate \\
\midrule
Bonferroni sequential $\chi^2$ ($p_t \leq \alpha/t$) & $\PoisHzeroBonfRate$ \\
Pocock $\alpha$-spending & $\PoisHzeroPocockRate$ \\
O'Brien--Fleming $\alpha$-spending & $\PoisHzeroOBFRate$ \\
Fixed-sample $\chi^2$ (batch) & $\PoisHzeroFCRate$ \\
Batch Fourier projection & $\PoisHzeroBFRate$ \\
E-process & $\PoisHzeroEPRate$ \\
Morozov (RMS rule) & $\PoisHzeroMorozovRate$ \\
\bottomrule
\end{tabular}
\end{table}
The $\alpha/t$ thresholds sum to $\sum_{t=1}^{T}\alpha/t = \alpha H_T$, which exceeds $\alpha$, so the rule spends more than its nominal budget and is anti-conservative by construction, independently of any dependence in the statistics. Despite its name it is therefore not a valid Bonferroni allocation, which would require $\sum_t \alpha_t \leq \alpha$.
This anti-conservatism is then amplified by the loose early thresholds: with $\alpha/t$ at $t=1$ equal to $\alpha$ itself, every trajectory whose first squared standardised residual lies in the upper $5\%$ tail of $\chi^2(1)$ rejects immediately. The same boundary shapes the rejection-time distribution under $H_1$ on our problems: across the bump, linear, and three-step misspecifications, $17$--$29\%$ of Bonferroni's rejections occur at $t < 5$ and $19$--$51\%$ at $t < 20$. Because the empirical null type-I rate ($\PoisHzeroBonfRate$) is itself concentrated at small $t$, a sizeable fraction of Bonferroni's apparent stopping-time advantage in Table~\ref{tab:detection} reflects rejections admitted by the $\alpha/t$ boundary rather than evidence-driven structural detection.

Pocock applies a uniform threshold across the trajectory and is mildly conservative on these cumulative-$\chi^2$ statistics ($0.030$). OBF concentrates its spending late and is mildly anti-conservative ($0.070$). The e-process is anytime-valid by Ville's inequality, so its type-I rate is bounded by $\alpha$ at every stopping time; the empirical $\PoisHzeroEPRate$ reflects the inequality's slack rather than anti-conservatism. Stopping-time comparisons in Tables~\ref{tab:reassurance} and~\ref{tab:icepack} should be read with these empirical rates in mind.

\section{Full power-curve breakdown}\label{sec:si-full-power}

Table~\ref{tab:si-full-power} reports detection rates and stopping times for all seven diagnostics across all four misspecification types and ten $\lambda$ values, complementing the headline Table~\ref{tab:detection} (which is restricted to E-process, Bonferroni, and Morozov for visual clarity). Three patterns are visible. First, the sequential magnitude tests (Pocock, OBF) track Bonferroni's detection curve qualitatively but with the empirical calibration shown in Table~\ref{tab:calibration} (Pocock conservative, OBF mildly anti-conservative). Bonferroni's median stopping times in the transition regions are very small (e.g.\ piecewise $\lambda = 0.10$: median $\PoisPieceLamPoneBonfMed$; three-step $\lambda = 0.15$: median $\PoisThreeLamPoneFiveBonfMed$; bump $\lambda = 0.10$: median $\PoisBumpLamPoneBonfMed$), reflecting that the $\alpha/t$ boundary admits a substantial fraction of small-$t$ rejections; Pocock and OBF report substantially larger medians because their rejection times reflect when evidence actually accumulates rather than when the threshold is loosest. Second, fixed $\chi^2$ and batch Fourier detect with equal or higher power than any sequential test in cells where the model error is super-noise (RMS$/\sigma > 1$), using the full $T = 501$ observations. Third, at the smallest linear signals the batch Fourier projection test is strongest and the e-process is the only sequential method with nontrivial structure-sensitive power. This is consistent with \S\ref{sec:linear-subnoise}: magnitude-based tests cannot resolve structured errors well below the noise, but recover power once the accumulated noncentrality grows.

The empirical null type-I rates differ across tests (Table~\ref{tab:calibration}) and should be kept in mind when comparing the columns: a meaningful fraction of Bonferroni's rejections occur at small $t$ where the $\alpha/t$ boundary is least stringent (Appendix~\ref{sec:si-calibration}), so its apparent speed advantage on three-step at low $\lambda$ is partly an artefact of the calibration scheme rather than evidence-driven detection.


\begin{table}[htbp]
\centering
\caption{Full power-curve breakdown across all four misspecification types and ten $\lambda$ values (100 Monte Carlo runs per cell, default bank $K = 156$). For each cell: detection rate for each diagnostic; median stopping time among detecting runs for the four sequential tests. Empirical null type-I rates: E-process $\PoisHzeroEPRate$, Bonferroni $\PoisHzeroBonfRate$, Pocock $\PoisHzeroPocockRate$, OBF $\PoisHzeroOBFRate$, fixed $\chi^2$ $\PoisHzeroFCRate$, batch Fourier $\PoisHzeroBFRate$, Morozov $\PoisHzeroMorozovRate$ (Table~\ref{tab:calibration}). For cells with detection rate below $0.20$, median stop times are computed over few detecting runs and should be read as illustrative. Bonferroni median stops of $1$--$2$ in low-detection cells reflect the $\alpha/t$ boundary, which is least stringent at $t = 1$ (SI~\S\ref{sec:si-calibration}).}
\label{tab:si-full-power}
\small
\begin{tabular}{@{}l c | c c c c c c c | c c c c@{}}
\toprule
& & \multicolumn{7}{c|}{Detection rate} & \multicolumn{4}{c}{Median stop} \\
\cmidrule(lr){3-9}\cmidrule(lr){10-13}
Type / $\lambda$ & RMS & E & Bonf & Pocock & OBF & FC$\chi^2$ & BF & Mrzv & E & Bonf & Pocock & OBF \\
\midrule
\multicolumn{13}{l}{\textit{Piecewise constant}} \\
\quad $\lambda = 0.02$ & 0.0010 & 0.07 & 0.07 & 0.03 & 0.06 & 0.03 & 0.30 & 0.00 & 227 & 1 & 23 & 369 \\
\quad $\lambda = 0.05$ & 0.0023 & 0.61 & 0.10 & 0.03 & 0.19 & 0.12 & 0.96 & 0.00 & 299 & 1 & 77 & 420 \\
\quad $\lambda = 0.08$ & 0.0035 & 0.99 & 0.19 & 0.28 & 0.62 & 0.55 & 1.0 & 0.00 & 165 & 1 & 278 & 351.5 \\
\quad $\lambda = 0.10$ & 0.0043 & 1.0 & 0.39 & 0.63 & 0.89 & 0.87 & 1.0 & 0.00 & 96.5 & 18 & 199 & 277 \\
\quad $\lambda = 0.15$ & 0.0060 & 1.0 & 0.93 & 1.0 & 1.0 & 1.0 & 1.0 & 0.00 & 48.5 & 135 & 122.5 & 177 \\
\quad $\lambda = 0.20$ & 0.0076 & 1.0 & 1.0 & 1.0 & 1.0 & 1.0 & 1.0 & 0.00 & 32.5 & 55 & 53 & 123 \\
\quad $\lambda = 0.30$ & 0.0102 & 1.0 & 1.0 & 1.0 & 1.0 & 1.0 & 1.0 & 0.00 & 18 & 11.5 & 19.5 & 75 \\
\quad $\lambda = 0.50$ & 0.0140 & 1.0 & 1.0 & 1.0 & 1.0 & 1.0 & 1.0 & 1.0 & 7.5 & 4 & 7 & 47 \\
\quad $\lambda = 1.00$ & 0.0194 & 1.0 & 1.0 & 1.0 & 1.0 & 1.0 & 1.0 & 1.0 & 4 & 3 & 4 & 29 \\
\addlinespace
\multicolumn{13}{l}{\textit{Gaussian bump}} \\
\quad $\lambda = 0.02$ & 0.0020 & 0.39 & 0.13 & 0.08 & 0.19 & 0.17 & 0.86 & 0.00 & 194 & 2 & 58 & 411 \\
\quad $\lambda = 0.05$ & 0.0047 & 1.0 & 0.52 & 0.70 & 0.96 & 0.94 & 1.0 & 0.00 & 77 & 86 & 153 & 254.5 \\
\quad $\lambda = 0.08$ & 0.0071 & 1.0 & 1.0 & 1.0 & 1.0 & 1.0 & 1.0 & 0.00 & 32 & 78 & 73 & 134 \\
\quad $\lambda = 0.10$ & 0.0086 & 1.0 & 1.0 & 1.0 & 1.0 & 1.0 & 1.0 & 0.00 & 23 & 29.5 & 40 & 103 \\
\quad $\lambda = 0.15$ & 0.0118 & 1.0 & 1.0 & 1.0 & 1.0 & 1.0 & 1.0 & 0.86 & 11 & 7 & 12 & 61.5 \\
\quad $\lambda = 0.20$ & 0.0145 & 1.0 & 1.0 & 1.0 & 1.0 & 1.0 & 1.0 & 1.0 & 9 & 5 & 8 & 47 \\
\quad $\lambda = 0.30$ & 0.0189 & 1.0 & 1.0 & 1.0 & 1.0 & 1.0 & 1.0 & 1.0 & 4 & 2 & 4 & 30.5 \\
\quad $\lambda = 0.50$ & 0.0248 & 1.0 & 1.0 & 1.0 & 1.0 & 1.0 & 1.0 & 1.0 & 3 & 2 & 3 & 15.5 \\
\quad $\lambda = 1.00$ & 0.0368 & 1.0 & 1.0 & 1.0 & 1.0 & 1.0 & 1.0 & 1.0 & 2 & 1 & 2 & 7 \\
\addlinespace
\multicolumn{13}{l}{\textit{Three-step}} \\
\quad $\lambda = 0.02$ & 0.0014 & 0.12 & 0.11 & 0.04 & 0.05 & 0.05 & 0.59 & 0.00 & 248 & 1 & 56.5 & 394 \\
\quad $\lambda = 0.05$ & 0.0033 & 0.99 & 0.15 & 0.21 & 0.56 & 0.45 & 1.0 & 0.00 & 147 & 5 & 207 & 361.5 \\
\quad $\lambda = 0.08$ & 0.0051 & 1.0 & 0.68 & 0.92 & 0.99 & 0.98 & 1.0 & 0.00 & 50 & 172.5 & 188.5 & 227 \\
\quad $\lambda = 0.10$ & 0.0063 & 1.0 & 0.98 & 0.99 & 1.0 & 1.0 & 1.0 & 0.00 & 34.5 & 135.5 & 109 & 164 \\
\quad $\lambda = 0.15$ & 0.0089 & 1.0 & 1.0 & 1.0 & 1.0 & 1.0 & 1.0 & 0.00 & 18.5 & 20.5 & 33 & 94.5 \\
\quad $\lambda = 0.20$ & 0.0114 & 1.0 & 1.0 & 1.0 & 1.0 & 1.0 & 1.0 & 0.62 & 11 & 7 & 14 & 64 \\
\quad $\lambda = 0.30$ & 0.0157 & 1.0 & 1.0 & 1.0 & 1.0 & 1.0 & 1.0 & 1.0 & 7 & 3 & 7 & 40.5 \\
\quad $\lambda = 0.50$ & 0.0227 & 1.0 & 1.0 & 1.0 & 1.0 & 1.0 & 1.0 & 1.0 & 3 & 2 & 3 & 19 \\
\quad $\lambda = 1.00$ & 0.0368 & 1.0 & 1.0 & 1.0 & 1.0 & 1.0 & 1.0 & 1.0 & 2 & 1 & 2 & 7 \\
\addlinespace
\multicolumn{13}{l}{\textit{Linear}} \\
\quad $\lambda = 0.02$ & 0.0003 & 0.01 & 0.07 & 0.07 & 0.07 & 0.05 & 0.03 & 0.00 & 31 & 1 & 52 & 415 \\
\quad $\lambda = 0.05$ & 0.0007 & 0.02 & 0.05 & 0.01 & 0.08 & 0.04 & 0.11 & 0.00 & 131.5 & 1 & 59 & 436.5 \\
\quad $\lambda = 0.08$ & 0.0011 & 0.03 & 0.08 & 0.03 & 0.10 & 0.04 & 0.31 & 0.00 & 171 & 1 & 65 & 393 \\
\quad $\lambda = 0.10$ & 0.0013 & 0.10 & 0.10 & 0.03 & 0.09 & 0.07 & 0.46 & 0.00 & 185 & 1 & 21 & 416 \\
\quad $\lambda = 0.15$ & 0.0019 & 0.22 & 0.13 & 0.04 & 0.15 & 0.12 & 0.84 & 0.00 & 182.5 & 1 & 147.5 & 409 \\
\quad $\lambda = 0.20$ & 0.0023 & 0.58 & 0.12 & 0.09 & 0.25 & 0.17 & 0.97 & 0.00 & 219.5 & 2 & 40 & 367 \\
\quad $\lambda = 0.30$ & 0.0031 & 0.97 & 0.16 & 0.18 & 0.47 & 0.39 & 1.0 & 0.00 & 186 & 2 & 275 & 358 \\
\quad $\lambda = 0.50$ & 0.0043 & 1.0 & 0.42 & 0.67 & 0.89 & 0.86 & 1.0 & 0.00 & 89.5 & 104 & 199 & 273 \\
\quad $\lambda = 1.00$ & 0.0057 & 1.0 & 0.87 & 0.97 & 1.0 & 1.0 & 1.0 & 0.00 & 50 & 148 & 160 & 201.5 \\
\bottomrule
\end{tabular}
\end{table}


\end{document}